\documentclass[aps,pra,twocolumn,showpacs,superscriptaddress]{revtex4}
\pdfoutput=1
\usepackage{graphicx}
\usepackage{amssymb}

\begin{document}

\title{Nonequilibrium Phase Diagram of a Driven-Dissipative Many-Body System}

\author{Andrea Tomadin}
\affiliation{Institute for Quantum Optics and Quantum Information of the Austrian Academy of Sciences, A-6020 Innsbruck, Austria}

\author{Sebastian Diehl}
\affiliation{Institute for Quantum Optics and Quantum Information of the Austrian Academy of Sciences, A-6020 Innsbruck, Austria}
\affiliation{Institute for Theoretical Physics, University of Innsbruck, Technikerstr. 25, A-6020 Innsbruck, Austria}

\author{Peter Zoller}
\affiliation{Institute for Quantum Optics and Quantum Information of the Austrian Academy of Sciences, A-6020 Innsbruck, Austria}
\affiliation{Institute for Theoretical Physics, University of Innsbruck, Technikerstr. 25, A-6020 Innsbruck, Austria}

\begin{abstract}
We study the nonequilibrium dynamics of a many-body bosonic system on a lattice, subject to driving and dissipation.
The time-evolution is described by a master equation, which we treat within a generalized Gutzwiller mean field approximation for density matrices.
The dissipative processes are engineered such that the system, in the absence of interaction between the bosons, is driven into a homogeneous steady state with off-diagonal long range order.
We investigate how the coherent interaction affects qualitatively the properties of the steady state of the system and derive a nonequilibrium phase diagram featuring a phase transition into a steady state without long range order.
The phase diagram exhibits also an extended domain where an instability of the homogeneous steady state gives rise to a persistent density pattern with spontaneously broken translational symmetry.
In the limit of small particle density, we provide a precise analytical description of the time-evolution during the instability.
Moreover, we investigate the transient following a quantum quench of the dissipative processes and we elucidate the prominent role played by collective topological variables in this regime.
\end{abstract}


\maketitle

\section{Introduction}
\label{sec:introduction}

One major challenge in nonequilibrium many-body physics is to identify situations with a sufficient degree of universality, i.e.~phenomena that occur independently of a precise microscopic realization and are largely insensitive to the specific choice of initial conditions. 
Several situations have been discussed in the literature, where many-body systems evolve in time towards nonequilibrium steady states.
The most prominent example in condensed matter is certainly the electron gas exposed to a bias voltage \cite{KL2009}.
In this context, truly many-body properties, such as the effect of nonequilibrium conditions on quantum critical points, have been investigated \cite{MTKM2006}.
Further implementations of nonequilibrium many-body systems are discussed in the context of exciton-polariton Bose-Einstein condensates \cite{ExcitonPolariton}, or more recently using driven noisy systems of trapped ions or dipolar atomic gases \cite{TDGA2010}.
As far as the time evolution of many-body systems is concerned, for closed systems we have seen a plethora of studies of quench dynamics \cite{QuenchTh,QuenchExp}, thermalization \cite{ThermalizationTh,ThermalizationExp}, and pre-thermalization \cite{BBW2004}, as well as transport \cite{Transport}.
Further dynamical studies involve situations of crossing quantum critical points in a finite time, in the spirit of the Kibble-Zurek mechanism \cite{KibbleZurekTh,KibbleZurekExp}, or investigations of the many-body Landau-Zener effect \cite{AG2008}. 

The quantum optics toolbox for the manipulation of cold atomic systems \cite{JZ2005} not only offers the possibility to tailor the coherent dynamics of a closed many-body system by Hamiltonian engineering, but also to implement the dissipative dynamics generated by a Liouville operator \cite{DMKKBZ2008}. 
Of particular interest are Liouville operators characterized by the existence of a unique dark state, i.e.~a sink state whereto the system is asymptotically driven, independently of the initial configuration.
In this case, it has been shown theoretically that the action of the sole dissipation allows to reliably prepare the system in peculiar states with interesting quantum mechanical features, such as phase coherence in bosonic systems \cite{DMKKBZ2008}, entanglement in spin systems \cite{VWC2009,WMLZB2009}, or delocalized pairing with arbitrary symmetry in fermionic systems \cite{DYDZ2010}.
Since a dark state can be reached by dissipative dynamics but not left, such processes featuring a dark states clearly violate the principle of detailed balance and hence may drive the system into a steady state quite different from thermodynamic equilibrium.
The existence of a dark state with known properties nevertheless promises a sufficiently universal nonequilibrium situation with a well-defined steady state.

Furthermore, since dissipation needs not be a small perturbation, but on the contrary can be made the dominant contribution to the dynamics, it is natural to investigate the effects of the interplay between unitary and dissipative dynamics.
In thermodynamic equilibrium, the competition of two non-commuting microscopic operators leads to a phase transition if the ground states have different symmetries when one or the other operator dominates \cite{SachdevBook}. 
A seminal example in the context of cold atoms in optical lattices is the superfluid-Mott insulator transition in the Bose-Hubbard model \cite{GMEHB2002}. 
By analogy with equilibrium physics, one may similarly expect a phase transition in the steady state of the system, resulting from the competition of unitary and dissipative dynamics. 

In this paper, we demonstrate a phase transition in the steady state of a driven-dissipative bosonic system on a lattice and discuss its properties in detail.
We point out several features of the phase transition that make it a genuine nonequilibrium phenomenon, not fitting into the well-developed theory of equilibrium quantum phase transitions.
However, we recognize some properties that may hold universally, i.e.~that may be shared by entire classes of similar systems, in the nonequilibrium steady state or in the time evolution.
The profound difference between the phenomena that we consider here and the equilibrium quantum phase transition manifests itself, at a purely technical level, in the absence of a known variational principle for the steady state.
The lack of such formal tool makes it necessary to evaluate the real-time evolution of the system, which is rather impractical in the many-body case.
Here we develop a mean field method to treat bosonic lattice systems, subject to unitary and dissipative dynamics, which generalizes the Gutzwiller ansatz and allows us to describe mixed states.
Within this approximation scheme, we deal with an effective master equation that acts on a reduced Hilbert space, is nonlinear in the quantum state, and is amenable of a workable numerical solution.
Moreover, we identify an analytically tractable low-density limit where the most relevant information on the quantum state is encoded in a few correlation functions, which obey coupled equations of motion.
The solution of the equations of motion, which we provide for various dynamical regimes in this approximation, explains qualitatively the behavior of the system at arbitrary density.
The method may also be used for semi-quantitative estimates in other problems where unitary and dissipative dynamics appear on equal footing. 

The paper is organized as follows.
In Sec.~\ref{sec:themodel} we introduce the microscopic model and provide a heuristic argument to justify the existence of a nonequilibrium phase transition in the model.
In Sec.~\ref{sec:meanfield} we introduce the generalized Gutzwiller mean field approximation, discuss the most prominent properties of the homogeneous phase diagram resulting from it, and point out profound differences with respect to equilibrium phase transitions in dissipative environments \cite{JJarrays}.
Sec.~\ref{sec:smalldens} specifies the Gutzwiller approximation to the case of low particle density and complements the numerical findings of Sec.~\ref{sec:meanfield} with analytical results.
Sec.~\ref{sec:instability} is devoted to the analysis of a dynamical instability that arises spontaneously in the nonequilibrium phase diagram. 
This completes the discussion of the steady state of the driven dissipative system. 
In Sec.~\ref{sec:revival} we focus on aspects of nonlinear dissipative dynamics that occur in mesoscopic settings. 
We draw our conclusions in Sec.~\ref{sec:conclusions}. 
The appendices \ref{app:meanfield}--\ref{app:eqmotion} provide technical details on the derivation of the mean field theory and on the numerical procedures used to solve the resulting nonlinear equations.

This paper refines the findings that we already presented in Ref.~\cite{DTMFZ2010}.
Here we provide a more detailed presentation of the material, including the discussion of the complete equations of motion and the explanation of numerical techniques.
New results include the exact solution of the dynamics for a small lattice (Sec.~\ref{sec:themodel}), the solution of the dynamically unstable equations of motion beyond linear response (Sec.~\ref{ssec:nonlineresponse}), as well as the investigation of the nonlinear dissipative dynamics following phase quenches (Sec.~\ref{sec:revival}).

\section{The model for competing unitary and dissipative dynamics}
\label{sec:themodel}

\subsection{ The model} 
\label{ssec:themodel}

In this work we consider a many-body system described by a master equation
\begin{equation}\label{eq:masterequation}
\partial_{t}\rho(t) = -i[\hat{\cal H},\rho(t)] + {\cal L}[\rho(t)]~.
\end{equation}
The master equation defines the time evolution of the density matrix $\rho(t)$ of an ensemble of bosonic atoms in an optical lattice.
It consists of a unitary part, described by the Hamiltonian $\hat{\cal H}$, and a dissipative part, represented by the Liouvillian ${\cal L}$. 
The Liouville operator arises from the coupling of the system to a bath, which is eliminated in the Markov approximation to yield the effective evolution (\ref{eq:masterequation}).
The validity of the Markov approximation, which neglects retardation effects in the bath, is rooted in the possibility of having strong separation of time scales for such quantum optical systems. 
We emphasize that this gives rise to a microscopic model which is \emph{local in time}. 
Therewith, the possibility of an accurate microscopic modeling for many-body systems in terms of a few parameters -- one of the key features of cold atomic many-body systems -- is extended from the Hamilton operator to the dissipative dynamics as well.
The unitary dynamics for the bosonic atoms in a $d$-dimensional lattice is described by the familiar Bose-Hubbard Hamiltonian
\begin{equation}\label{eq:bosehubbard}
\hat{\cal H} = - J \sum_{\langle \ell\ell'\rangle} \hat{b}_{\ell}^{\dag}\hat{b}_{\ell'} + \frac{U}{2}\sum_{\ell}\hat{n}_{\ell}(\hat{n}_{\ell}-1) -\sum_{\ell}\mu\hat{n}_{\ell}~,
\end{equation}
where $J$ is the hopping amplitude, $U > 0$ the interaction strength, and $\mu$ the chemical potential (see Sec.~\ref{ssec:chempot} and App.~\ref{app:lingeneral} for a discussion of its role in the time-evolution of the system).
The operator $\hat{b}_{\ell}$ ($\hat{b}_{\ell}^{\dag}$) annihilates (creates) a boson in the $\ell$th lattice site and $\langle \ell \ell' \rangle$ indicates that the summation is restricted to nearest neighboring sites.
($\ell$ is a vector of integers with the same dimension as the lattice.)
The Liouvillian ${\cal L}$ that we consider here was derived in Ref.~\cite{DMKKBZ2008}.
Written in Lindblad form, it reads
\begin{equation}\label{eq:liouvillian}
{\cal L}[\rho(t)] = \frac{1}{2}\kappa\, \sum_{\langle \ell\ell'\rangle} [2\hat{c}_{\ell\ell'}\rho \hat{c}_{\ell\ell'}^{\dag} - \hat{c}_{\ell\ell'}^{\dag}\hat{c}_{\ell\ell'}\rho-\rho \hat{c}_{\ell\ell'}^{\dag} \hat{c}_{\ell\ell'}]~,
\end{equation}
where $\kappa$ is the dissipation rate and the quantum jump operators $\hat{c}_{\ell,\ell'}$, acting on pairs of neighboring lattice sites, are given by
\begin{eqnarray}\label{eq:jumpoperators}
\hat{c}_{\ell\ell'}
& = & (e^{-i \phi_{\ell}} b_{\ell}^{\dag} + e^{-i \phi_{\ell'}} b_{\ell'}^{\dag}) (e^{i \phi_{\ell}} b_{\ell} - e^{i \phi_{\ell'}} b_{\ell'}) \nonumber \\
& = & (b_{\ell}^{\dag}+e^{-i\phi_{\ell\ell'}}b_{\ell'}^{\dag}) (b_{\ell}-e^{i\phi_{\ell\ell'}}b_{\ell'})~.
\end{eqnarray}
with $\phi_{\ell\ell'} = \phi_{\ell'} - \phi_{\ell}$.
We specialize to the case that the phase difference along the $\lambda = 1 \dots d$ primitive directions ${\bf e}_{\lambda}$ of the lattice is a constant $\phi_{\lambda} = \phi_{\ell + {\bf e}_{\lambda}} - \phi_{\ell}$.
The quantum jump operators are composed of an annihilation part, which destroys a particular superposition of the bosonic wavefunction, and a creation part, which recycles the bosons after interacting with the bath. 
While the choice of the phase in the annihilation part is crucial to the asymptotic behavior of the system (see below), the phase in the creation part is a matter of convention and could also be omitted.
One possible implementation of these jump operators in optical superlattices is discussed in Ref.~\cite{DMKKBZ2008}. 
The key point of such choice for the jump operators is that their nullspace consists of a unique element $| D\rangle $ for a fixed number of particles $N$.
\begin{figure}[t]
\begin{center}
\includegraphics[width=1.00\linewidth]{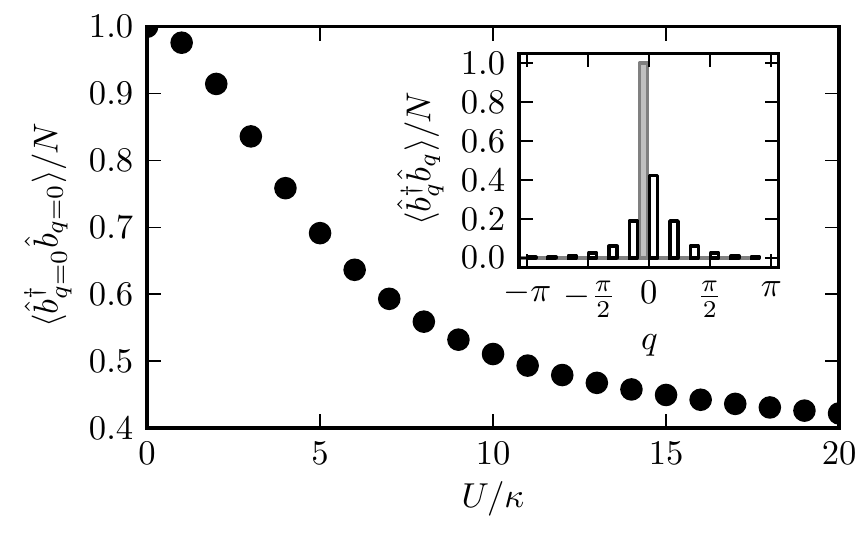}
\caption{The main panel shows the $q=0$ entry of the single-particle correlation matrix in momentum space as the interaction strength is varied, obtained from the solution of the exact EOMs (\ref{eq:masterequation}) for $N=2$, $L=12$, $J=0.0$, and $\phi=0$, at the time $T_{\rm f} \kappa = 20.0$.  
The inset shows the entire diagonal of the correlation matrix for $U/\kappa = 0.0$ (shaded bars) and $U/\kappa = 20.0$ (empty bars).}
\label{fig:exactcorr}
\end{center}
\end{figure}
It follows that $| D \rangle$ is the unique dark state for the whole Liouville operator and that the system is attracted to this state independently of the initial conditions, if no competition arises from the unitary part of the evolution. 
In Refs.~\cite{DMKKBZ2008,KBDKMZ2009} it was shown that the state
\begin{eqnarray}\label{eq:becstate}
|{\rm BEC}_{{\bf q}} \rangle = (N!)^{-1/2} \left ( \sum_{\ell} e^{-i {\bf q} \ell} \hat{b}_{\ell}^{\dag} \right ) 
= (N!)^{-1/2} b_{{\bf q}}^{\dagger\, N}|{\rm vac}\rangle
\end{eqnarray}
[a Bose-Einstein condensate (BEC) with momentum ${\bf q} = - \phi_{\lambda} {\bf e}_{\lambda}$ in lattice units] is the unique dark state of the Liouvillian (\ref{eq:liouvillian}).
The uniqueness can be easily understood in momentum space, where the annihilation part of $\hat{c}_{\ell\ell'}$ reads $\sum_{\lambda} \{1 - \exp{[- i (q_{\lambda} + \phi_{\lambda})} ]\} \hat{b}_{\bf q}$.
Since the creation part of $\hat{c}_{\ell,\ell'}$ does not have any eigenvalue on the Hilbert space of $N-1$ particles, $\hat{c}_{\ell\ell'}$ features a unique dissipative zero mode at the given momentum.
The existence of a dark state depends here on the jump operators being not hermitian.
It is important to remark, anyway, that the Liouville operator conserves the total particle number $\hat{N} = \sum_{\ell} \hat{n}_{\ell}$, as the jump operators just redistribute a particle's superposition and thus fulfill $\exp{ ( + i \phi \hat{N} ) } \hat{c}_{\ell\ell'} \exp{ ( - i \phi \hat{N} ) } = \hat{c}_{\ell\ell'}$.
However, the non-hermitian nature of the jump operators entails that the dissipative process represented by the Liouvillian (\ref{eq:liouvillian}) does not respect the principle of detailed balance, because the probability that the system is projected out of the dark state is zero.
The dissipative dynamics produced by the engineered Liouvillian is necessarily very different from a thermalization process and hence the phase portrait of the system cannot be understood in terms of a modified or perturbed equilibrium picture, but requires a genuinely new approach.

An alternative physical picture of the dynamics in the presence of the dark state is the following. 
On two neighboring sites along the $\lambda$ direction, the component of the wave function that is not in the nullspace of $\hat{b}_{\ell} - e^{i \phi_{\lambda}} \hat{b}_{\ell + {\bf e}_{\lambda}}$ is mapped onto the whole Hilbert space, including the nullspace, while the nullspace is mapped onto itself.
Asymptotically, the weight of the wave function outside the nullspace vanishes and the phase of the wave function between the neighboring sites is locked to $-\phi_{\lambda}$.
 Since the phase locking takes place on each pair of sites, the many-body state is asymptotically projected into the BEC state, i.e.~a state with long-range phase coherence.
The dissipative evolution purifies the initial state of the system and, in this sense, pumping into the BEC bears strong analogies to laser cooling of single atoms \cite{LaserCooling}. 
The analogy to optical pumping is quite tight, because the implementation of the dissipative process involves a coherent (laser) driving element, such that the system is indeed driven-dissipative.

\subsection{Competition between unitary and dissipative dynamics}
\label{ssec:competition}
\begin{figure}[t]
\begin{center}
\includegraphics[width=1.00\linewidth]{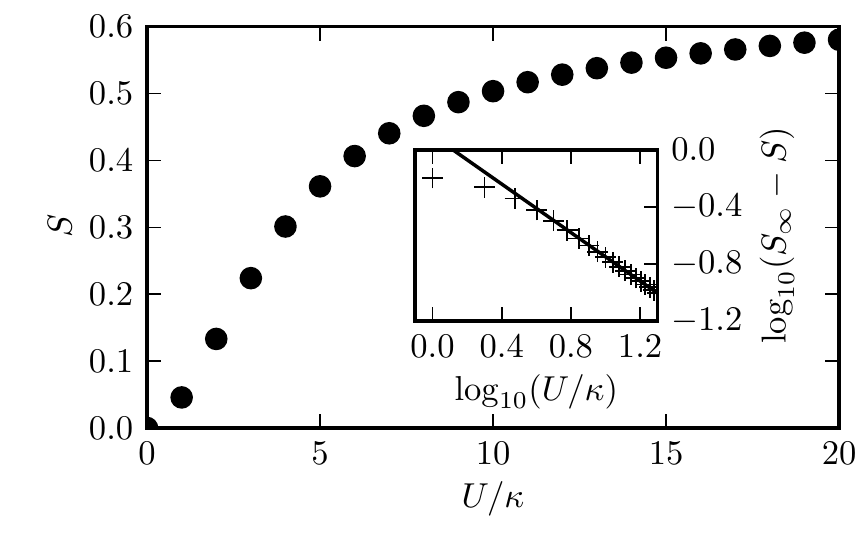}
\caption{The main panel shows the entropy in the steady state of the system, from the solution of the exact EOMs (\ref{eq:masterequation}) for the same parameters as Fig.~\ref{fig:exactcorr}.  
The inset shows the same data as the main panel (crosses) logarithmically rescaled with respect to the asymptotic value $S_{\infty} \simeq 0.72$ and compared to the power-law $S_{\infty}-S \propto U^{-\alpha}$ for $\alpha \lesssim 1.0$ (solid line).}
\label{fig:exactentropy}
\end{center}
\end{figure}

We start out by considering the effect of strong interactions on the steady state of the system.
In the equilibrium phase diagram of the Bose-Einstein Hamiltonian~(\ref{eq:bosehubbard}), the quantum phase transition between the Mott insulator and the superfluid is understood in terms of the competition between the kinetic term proportional to $J$, that tends to delocalize the bosons to decrease the energy, and the interaction term proportional to $U$, that is responsible for an increase of the energy when particles tunnel between sites and repel each other.
To understand the interplay of the energy scale in our present setup, a novel point of view is required, because the time-evolution of the system does not minimize the energy.
The absence of a variational principle to characterize the steady state of the system is indeed a noteworthy difference with respect to the equilibrium theory.
Still, we identify the existence of a competition between the energy scales of the interaction $U$ and the dissipation $\kappa$.
To this end, we proceed by adopting a rotating frame that eliminates the interaction from the Hamiltonian.
The local operator that defines the new frame on each site is
$\hat{V}_{\ell} = \exp{[i U \hat{n}_{\ell} (\hat{n}_{\ell} - 1) t]}$.
The annihilation operator in the new frame reads
\begin{equation}
\hat{V} \hat{b}_{\ell} \hat{V}^{-1} 
= \exp{[- i U \hat{n} t ]} \hat{b}_{\ell} 
= \sum_{n} \exp{[ i n U t]} |n\rangle_{\ell}\langle n| \hat{b}_{\ell}~.
\end{equation}
The effect of the interaction is to rotate the phase of each Fock state differently, destroying the phase coherence of the order parameter.
Hence we argue that increasing the interaction leads to a substantial depletion of the condensate and eventually to the disappearance of the long-range order in the system.

Before developing a suitable mean field approximation that allows us to tackle the question of the  behavior of the system in the thermodynamic limit, we provide signatures of the competition between interaction and dissipation from exact numerical evaluation in a small system.
For this purpose, we integrate numerically the equations of motion (EOMs)~(\ref{eq:masterequation}), in the Hilbert space for $N=2$ particles in $L = 12$ lattice sites with periodic boundary conditions, using a fixed-stepsize fourth-order Runge-Kutta method.
We use the final density matrix $\rho (T_{\rm f})$ to compute the single-particle correlation matrix $\langle \hat{b}_{q}^{\dag} \hat{b}_{q'} \rangle$ and the entropy $S = - {\rm Tr}[ \rho \ln{\rho} ]$, where $q$,$q'$ are momenta in the discrete Brillouin zone of the finite lattice.
Fig.~\ref{fig:exactcorr} shows that, as the interaction strength is increased, the occupation of the zero-momentum state decreases monotonically. 
This is a signature, in the finite system, of the depletion of the condensate fraction that we expect in the thermodynamic limit.
We also see that the distribution of the momenta is localized in the absence of interactions and broadens substantially for $U \gtrsim \kappa$, analogously to the broadening of the condensate peak produced by a finite temperature.
Such broadening is not due to a rearrangement of the single-particle states, e.g.~the momentum states being not the appropriate eigenbasis for the system, because, as we show in Fig.~\ref{fig:exactentropy}, the entropy steadily increases with the interaction strength.
This means that, in the presence of interactions, the Liouvillian is no longer able to drive the system towards the pure dark state.
It is interesting to note that the entropy converges towards a definite constant value as the interaction strength grows, as is clear from the analysis in the inset of Fig.~\ref{fig:exactentropy}.
This fact suggests that the crossover exhibited by the finite system may in fact turn into a transition from the condensed state to a well-defined new phase.
In the following section we validate this picture and provide a detailed analysis of the phase diagram.

\section{Nonequilibrium phase diagram in the mean field approximation}
\label{sec:meanfield}

In order to capture the physics of the above depicted nonequilibrium phase transition on a semi-quantitative level, we develop a formalism which allows us to describe the physics of both the well-controlled limits of weak and strong interactions.
In the weakly interacting limit $U/\kappa\ll1$ we must recover the condensate phase with long-range order found in Ref.~\cite{DMKKBZ2008}, while in the strong coupling limit  $U/\kappa \gg 1$ we expect a disordered phase with diagonal density matrix.
To this end, we tackle the solution of Eq.~(\ref{eq:masterequation}) within a mean field approximation, in the form of a Gutzwiller-like ansatz where the system density matrix is factorized in position space
\begin{equation}\label{eq:factorization}
\rho = \bigotimes_{\ell}\rho_{\ell}~.
\end{equation}
The reduced density matrix $\rho_{\ell} = {\rm Tr}_{\neq \ell}\rho$ on the site $\ell$ is obtained by tracing out all the the other degrees of freedom but those of the $\ell$th site.
The equation of motion for the reduced density matrix then reads
\begin{equation}\label{eq:motiontrace}
\partial_{t}\rho_{\ell}(t) = -i {\rm Tr}_{\neq \ell}[\hat{\cal H},\rho(t)] + {\rm Tr}_{\neq \ell}{\cal L}[\rho(t)]~.
\end{equation}
The computation of the traces is reported in App.~\ref{app:meanfield}.
The Hamiltonian part results in a local commutator
\begin{equation}
-i {\rm Tr}_{\neq \ell}[\hat{\cal H},\rho(t)] = -i[\hat{h}_{\ell}(t), \rho_{\ell}(t)]~,
\end{equation}
with the effective reduced local Hamiltonian given by $\hat h_\ell = -J\sum_{\langle \ell'| \ell\rangle}(\langle \hat{b}_{\ell'} \rangle \hat{b}_{\ell}^{\dag} +\langle \hat{b}_{\ell'}^{\dag} \rangle \hat{b}_{\ell}) + \frac{1}{2} U \hat{n}_{\ell} (\hat{n}_{\ell} - 1) - \mu \hat{n}_{\ell}$ [cf.~also Eq.~(\ref{eq:localhamiltonian})]. 
This is precisely the form obtained in the ordinary Gutzwiller approximation for a mean field ground state calculation on the Bose-Hubbard model.
Note, however, that due to the restriction of the ground state, the state must be pure and thus the onsite density matrices in the latter case are of the form $\rho_{\ell} = | \psi \rangle_{\ell} \langle \psi |$.
In our case we do not have an argument in favor of the purity of the steady state for strong interactions so we allow for mixed steady states as well.

The trace of the Liouvillian part, assuming an axial geometry in the system (see App.~\ref{app:meanfield}), produces the structure
\begin{eqnarray}\label{eq:liouvtrace}
{\rm Tr}_{\neq \ell}{\cal L}[\rho(t)] & = & \kappa \sum_{r,s} [2 A_{\ell}^{r} \rho_{\ell} A_{\ell}^{s \dag} - A_{\ell}^{s \dag} A_{\ell}^{r} \rho_{\ell} - \rho_{\ell} A_{\ell}^{s \dag} A_{\ell}^{r}] \nonumber \\
& \times & [\Gamma_{\ell, \sigma = +1}^{r s} + \Gamma_{\ell, \sigma = -1}^{r s} + (z-2) \Gamma_{\ell, \sigma = 0}^{r s}]~,	
\end{eqnarray}
where $z$ is the coordination number of the lattice, the vectors of operators ${\bf A}$ are given in Eq.~(\ref{eq:liouvdecomp}), and the matrix of coefficients $\Gamma_{\ell,\sigma}$ in Eq.~(\ref{eq:motionlattice}).

At this point we discuss the validity of the factorization approximation (\ref{eq:factorization}) in real space.
First we note that the dimensionality of the system enters the EOMs only through the coordination number $z$ of the lattice.
In general, one expects higher reliability of mean field approximations in higher spatial dimensions.
For the (fermionic) Hubbard model, arguments based on the central limit theorem for spatial dimension $d \to \infty$ lead to the observation that mean field theory becomes exact in this limit \cite{MV1989}. 
The statistical nature of the argument suggest that a similar reasoning could be applied also here. 
The final results of the analysis do not depend qualitatively on the precise value of $z$ hence, when the choice of a definite dimensionality is necessary, we consider for simplicity a one-dimensional ($d = 1$) system.
Moreover we observe that the system is represented only locally as a quantum state, i.e.~a vector or a density matrix in the local bosonic Fock state of each site $\ell$.
Any correlation function $\langle \hat{\cal O}_{\ell} \hat{\cal O}_{\ell'} \rangle$ that involves the evaluation of operators on different sites simplifies to the product $\langle \hat{\cal O}_{\ell} \rangle \langle \hat{\cal O}_{\ell'} \rangle$.
The connected part of the correlation function between different sites is entirely neglected.
However, the onsite correlations are evaluated exactly and thus this method is quite suitable to extract informations on a wide range of variation of $U / \kappa$, from the weak to the strong coupling limit. 
In the limit of weak coupling, additional arguments in favor of our approximation scheme are available (see Sec.~\ref{ssec:linresponse}).

Technically, the Gutzwiller approximation transforms the full many-body system into a collection of coupled single-site systems, that are amenable of a much simpler analytical treatment.
Moreover, in a numerical solution, the total dimension of the state vector of the system decreases from the full size of the Hilbert space to $L$ times the size of the local Hilbert space.
This makes much larger arrays numerically tractable.
On the technical side, the price to pay for the decrease in size of the Hilbert space is that the equation of motion becomes nonlinear in the state vector, contrary to the Schr\"odinger and von Neumann equations.
More precisely, the coefficients that enter the effective local Hamiltonian $\hat{h}_{\ell'}$ and the matrix $\Gamma_{\ell',\sigma}$ depend on a set of local correlation functions
\begin{equation}\label{eq:corrset}
\psi_{\ell} \equiv \langle \hat{b}_{\ell} \rangle, \quad \langle \hat{b}_{\ell}^{2} \rangle, \quad \langle \hat{n}_{\ell} \hat{b}_{\ell} \rangle~,
\end{equation}
evaluated with the density matrix of the sites $\ell$ neighbors of $\ell'$.
The appearance of a nonlinear equation from an originally linear equation for the system density operator in the framework of a mean field approximation is actually familiar from e.g.~the Gross-Pitaevski mean field theory for weakly interacting Bose gases. 
There, a nonlinear classical field equation for the condensate expectation value results from a mean field approximation on the original linear $N$-body Schr\"odinger equation. 
Finally, the mean field equations conserve the average value $n$ of the total particle density but do not commute with the number operator.
We reiterate that the treatment of the correlation functions (\ref{eq:corrset}) on each site is fully quantum mechanical, i.e.~there is no approximation on the value of the average of any local operator.
In particular, this procedure yields a theory that contains much more information than a Gross-Pitaevski equation [see Eq.~(\ref{eq:dissgpe}) and the following discussion], where also the product of local operators (\ref{eq:corrset}) is factorized, so that all available information is encoded in $\psi_{\ell} $.

The effect of the phase $\phi$ in the Liouvillian is easily understood for a homogeneous state by considering the change of frame of reference $\tilde{b}_{\ell} = e^{+ i \phi \ell} \hat{b}_{\ell}$.
In the new frame, the Liouvillian maintains the same form, with $\hat{b}_{\ell} \mapsto \tilde{b}_{\ell}$ and $\phi \mapsto 0$.
In the effective Hamiltonian, instead, it is necessary to apply $\hat{b}_{\ell} \mapsto \tilde{b}_{\ell}$ but also $J \mapsto J \cos{\phi}$.
Hence, we can reduce to considering the homogeneous states of the system with $\phi = 0$ only, as the general case amounts to a mere reduction of $J$ if $\phi < \pi / 2$.
For $\phi > \pi / 2$, however, the dark state of the Liouvillian corresponds to a boosted condensate that is known to be dynamically unstable \cite{APDHL2005} and consequently does not feature a stable homogeneous steady state.

\subsection{Phases of the homogeneous steady state.}
\label{ssec:phases}

In this section we study the steady states of Eq.~(\ref{eq:motiontrace}) and validate the picture, suggested in Sec.~\ref{ssec:competition}, of an interaction-driven nonequilibrium phase transition that destroys the condensation in the system.
We first identify analytically two exact steady states, for vanishing and large interactions, and then provide numerical results that interpolate between these two cases.

Let us consider first the pure coherent state \cite{GardinerZollerBook}
\begin{equation}\label{eq:cohestate}
|\psi\rangle_{\ell} = \sum_{\nu=0}^{\infty}e^{-|\psi_{\ell}|^{2}/2}\frac{\psi_{\ell}^{\nu}}{\sqrt{\nu!}}|\nu\rangle_{\ell}~,
\end{equation}
(where $\nu$ is the index in the local Fock space) and the corresponding density matrix $\rho^{\rm (c)} = | \psi \rangle_{\ell} \langle \psi|$.
The Liouvillian (\ref{eq:liouvtrace}) vanishes exactly when applied to the homogeneous coherent density matrix, that provides hence a consistent dark state solution for the dissipative dynamics and pinpoints an exact steady state solution of the model at $J = U = 0$.
Moreover, for this density matrix, all onsite correlation functions factorize in Fock space.
As a consequence, all the information of the theory is encoded in the simplest one, namely the one-point function $\langle \hat{b}_{\ell} \rangle$, which is precisely the order parameter (\ref{eq:corrset}) of the condensate.
The equation of motion $\partial_{t} \langle \hat{b}_{\ell} \rangle = {\rm Tr}[\hat{b}_{\ell} \partial_{t} \rho_{\ell}(t)]$ for the order parameter reads explicitly
\begin{eqnarray}\label{eq:dissgpe}
- i\partial_{t} \psi_{\ell}(t) & = &
J \sum_{\langle \ell' | \ell \rangle} \psi_{\ell'} +  \mu \psi_{\ell} -  U |\psi_{\ell}|^{2} \psi_{\ell} \nonumber \\
& & +2 i \kappa \sum_{\langle \ell' | \ell \rangle} (\psi_{\ell} - \psi_{\ell'} + \psi_{\ell'}^{\ast} \psi_{\ell}^{2} - |\psi_{\ell'}|^{2} \psi_{\ell} )~.
\end{eqnarray}
This equation constitutes a dissipative Gross-Pitaevski equation (GPE) and, indeed, the unitary part is identical to the standard lattice GPE.
If we choose a homogeneous state $\psi_{\ell} =\psi$ [analogous to the BEC state (\ref{eq:becstate})], the dissipative terms vanish and the equation reduces to $-i \partial_{t} \psi =  (z J - U \psi^{\ast} \psi + \mu) \psi$, i.e.~the system undergoes a global phase rotation that can be removed with a precise choice of the chemical potential $\mu = - z J + U n$.
More generally, we will constantly take advantage of the possibility to fix the chemical potential and remove residual phase rotations, ensuring a time-independent steady state as illustrated in this simple case. 
The role of the chemical potential in our nonequilibrium analysis is further discussed below in Sec.~\ref{ssec:chempot}. 
The coherent state is not a general solution of Eq.~(\ref{eq:motiontrace}) precisely because a different phase frequency appears for each different correlation functions but only one can be removed with the chemical potential.
Anyway, the similarity of the Eq.~(\ref{eq:dissgpe}) to the GPE suggests that small values of the interaction strength $U$ do not change qualitatively the dynamics of the system, which is the result rigorously derived in Ref.~\cite{DMKKBZ2008} within the Bogoliubov approximation.
\begin{figure}[t]
\begin{center}
\includegraphics[width=0.90\linewidth]{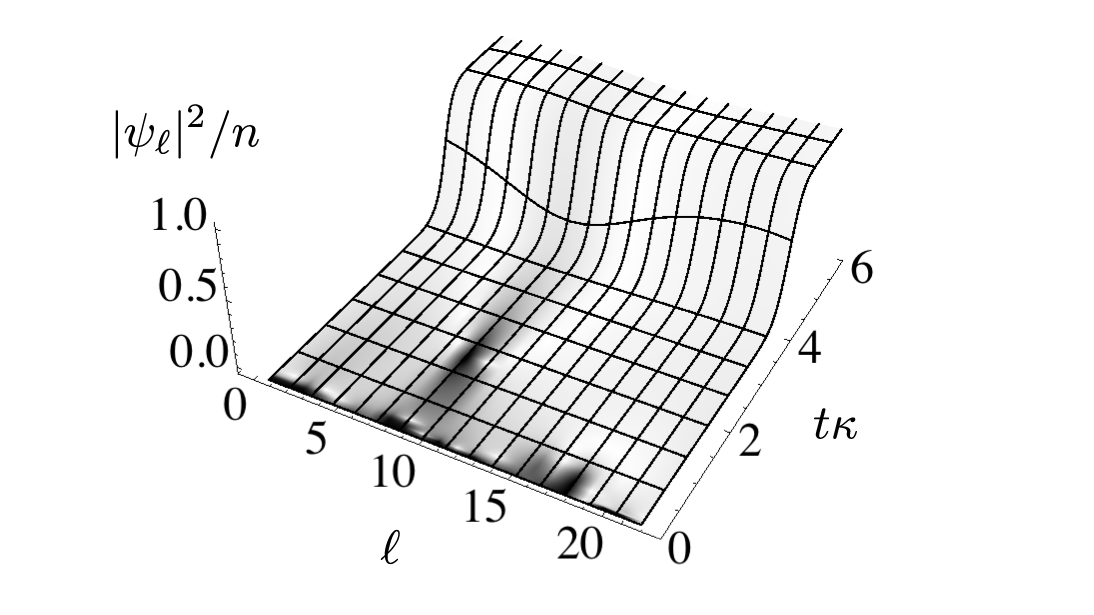}
\caption{Time evolution of the condensate fraction $|\psi_{\ell}|^{2}/n$ on the sites $1\le \ell \le L = 22$ of the lattice, according to the mean field EOM (\ref{eq:motiontrace}), for $n=1.0$, $J = 0.0$, $U = 3.0$, and $\phi = 0$.  
The color of the surface represents the phase difference $\delta_{\ell}$ between neighboring sites (white for $\delta_{\ell}=0$, black for $\delta_{\ell}=\pm \pi$).}
\label{fig:condensation3d}
\end{center}
\end{figure}

Let us consider now the opposite limit of a diagonal density matrix state, in which the coherences are totally absent.
According to the heuristic argument presented in Sec.~\ref{ssec:competition}, this situation is expected for large values of the interaction strength.
In this case, the kinetic term of the Hamiltonian drops out, since no off-diagonal order is present and thus $\langle \hat{b}_{\ell} \rangle=0$.
The interaction term drops out as well, because the operator $\hat{n}_{\ell}$ commutes with any state that is diagonal in the Fock basis.
The only remaining energy scale in the system is the dissipation strength $\kappa$.
Finally, the matrix of coefficients in Eq.~(\ref{eq:liouvtrace}) reads $\Gamma_{\ell,\sigma} = {\rm diag}(0, \langle \hat{n}_{\ell} \rangle, \langle \hat{n}_{\ell} \rangle + 1, 1)$ and the EOM for a homogeneous system reads
\begin{eqnarray}\label{eq:thevol}
\partial_{t} \rho_{\ell}(t) & = & 2 \kappa (n + 1)(2 \hat{b}_{\ell} \rho_{\ell} \hat{b}_{\ell}^{\dag} - \hat{b}_{\ell}^{\dag} \hat{b}_{\ell} \rho_{\ell} - \rho_{\ell} \hat{b}_{\ell}^{\dag} \hat{b}_{\ell}) \nonumber \\
& & + 2 \kappa n (2 \hat{b}_{\ell}^{\dag} \rho_{\ell} \hat{b}_{\ell} - \hat{b}_{\ell} \hat{b}_{\ell}^{\dag} \rho_{\ell} - \rho_{\ell} \hat{b}_{\ell} \hat{b}_{\ell}^{\dag})~.
\end{eqnarray}
The latter equation of motion describes precisely a bosonic mode $\hat{b}_{\ell}$ coupled to a bath with thermal occupation $n = \langle \hat n_\ell\rangle$ \cite{GardinerZollerBook}.
It is intriguing that in this case $n$ is a property of the system itself and hence we may say that the system acts as its own bath. In particular, this effective thermal bath, which drives the dynamics, is not related to the physical bath that is necessary to implement the jump operators~(\ref{eq:jumpoperators}).
In the steady state $\partial_{t} \rho_{\ell}=0$ the remaining energy (rate) scale $\kappa$ drops out, and the only physical scale is the average particle number $n$. The steady state solution is readily seen to be a thermal-like state
\begin{equation}\label{eq:thrmstate}
\rho_{\ell}^{\rm (th)} = \sum_{\nu=0}^{\infty} | \nu \rangle_{\ell} \frac{n^{\nu}}{(n+1)^{(\nu + 1)}} \langle \nu |_{\ell}~,
\end{equation}
We emphasize that this state emerges \emph{despite} the fact that the system is driven out of thermodynamic equilibrium, not as a consequence of the latter. 
\begin{figure}[t]
\begin{center}
\includegraphics[width=1.00\linewidth]{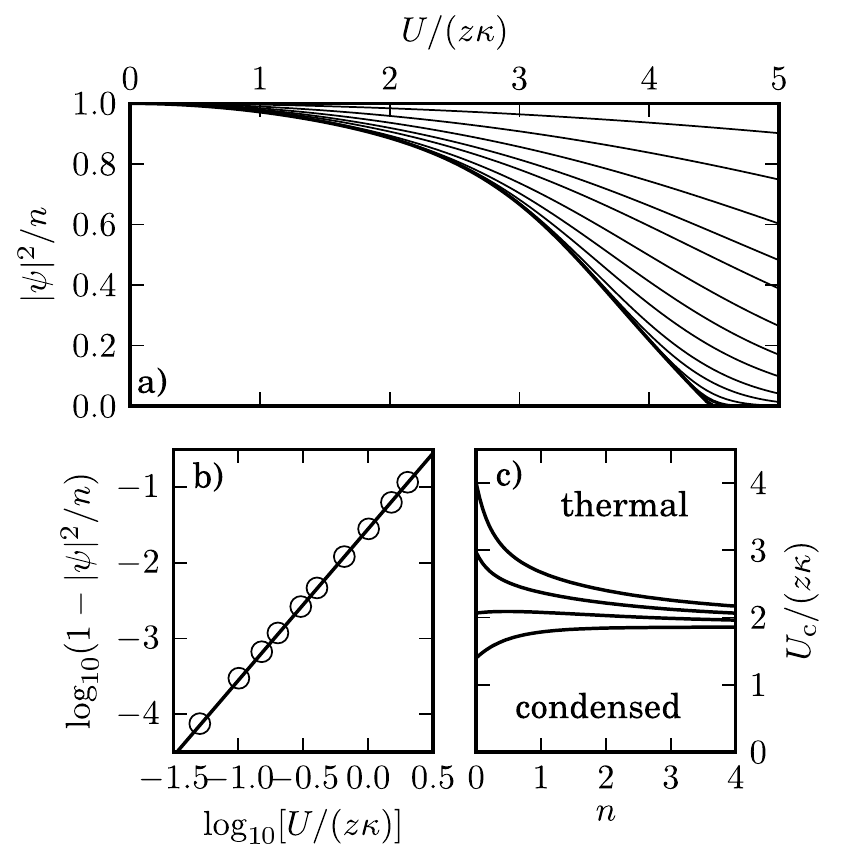}
\caption{
Characterization of the homogeneous steady state for $J = 1.5 \kappa$ and $\phi = 0$.
(a) Stroboscopic plot of the time evolution of the condensate fraction for $n = 1.0$, starting from an initially fully condensed state.
(b) Comparison of the full numerical solution of the homogeneous system for $n = 0.1$ to the analytical low density result.
(c) Critical point in the homogeneous phase diagram as a function of density for different values of $J = 0.0$, $0.5$, $1.0$, $1.5$.}
\label{fig:depletion} 
\end{center}
\end{figure}
Finally, we note that the nonlinearity of the physically motivated mean field approximation allows us to describe two distinct dynamical fixed points, namely one ordered state with finite off-diagonal expectation values such as $\langle \hat{b}_{\ell} \rangle$, and one disorder phase with vanishing coherences, in some analogy to $\phi^{4}$-theory for weakly interacting boson systems. 
We also emphasize that the breaking of the $U(1)$ phase symmetry (associated to the conservation of particle number) takes place spontaneously in the ordered phase, since both Hamiltonian and Liouville operator in the original equations (\ref{eq:masterequation}--\ref{eq:liouvillian}) conserve the number of particles.

We turn now to the numerical integration of the EOM~(\ref{eq:motiontrace}), specified to a lattice with $L$ sites and periodic boundary conditions.
The integration is performed with a fourth-order Runge-Kutta algorithm, using a fixed stepsize typically taken $\delta t = 10^{-3} \kappa^{-1}$ \cite{footnote0}.
The local Hilbert space is truncated to a maximum particle number $\nu_{\rm max} \lesssim 20$, substantially larger than the typical particle densities $n \lesssim 1.0$ that we consider here.
Due to the periodicity of the lattice, the momenta are confined to a discretized Brillouin zone such that the phase $\phi$ assumes the values $2 \pi m / L$, with $m \in \mathbb{N}$.
For each site $\ell$ in the lattice, the execution of the integration returns the local correlation functions.
To measure the degree of coherence along the array, we compute the phase differences $\delta_{\ell}$ between the first-order correlation functions on neighboring sites
\begin{equation}
2 \pi \delta_{\ell} / L = {\rm Im} \ln \left ( \frac{\langle \hat{b}_{\ell} \rangle \langle \hat{b}_{\ell + 1} \rangle^{\ast}} {| \langle \hat{b}_{\ell} \rangle \langle \hat{b}_{\ell + 1} \rangle|^{2}} \right )~.
\end{equation}
A state with definite momentum is such that $\delta_{\ell}$ is integer and constant along the array.
A typical output of the numerical integration is displayed in Fig.~\ref{fig:condensation3d}, where we show the condensate fraction $|\psi_{\ell}|^{2} / n$ as a function of time on a lattice with $L = 22$.
The array is initialized as a collection of thermal states (\ref{eq:thrmstate}), slightly perturbed with tiny random off-diagonal fluctuations.
The figure shows that the system is driven out of its initial thermal state, develops a uniform macroscopic condensate fraction $\lesssim 1.0$ and, at the same time, the fluctuations of the local correlations vanish and a definite phase difference $\delta_{\ell} = 0$ is established throughout the array.
The remarkable property of the Liouvillian (\ref{eq:liouvillian}), namely that local phase synchronization results in the establishing of long-range order~\cite{DMKKBZ2008}, is visually clear in this picture.
\begin{figure}[t]
\begin{center}
\includegraphics[width=1.00\linewidth]{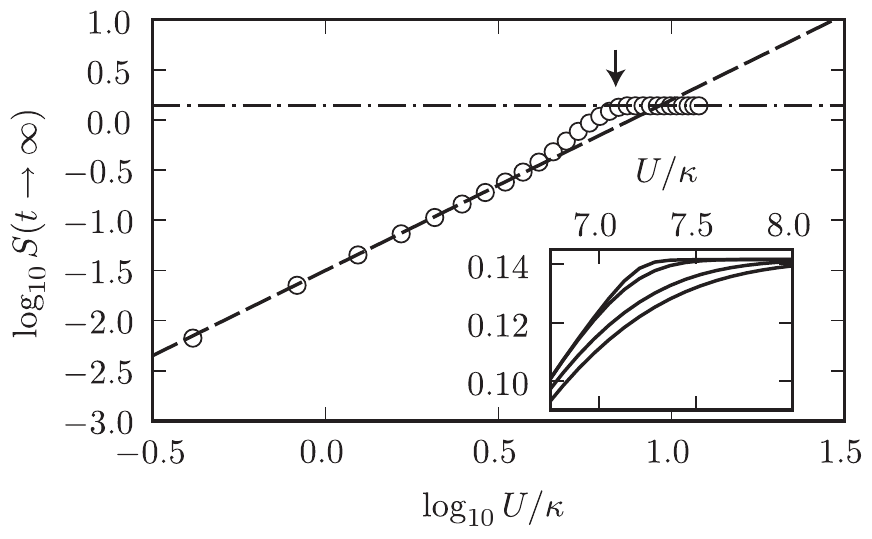}
\caption{Asymptotic value of the entropy $S$ as the interaction strength $U$ is increased, for $J = 0.0$, $\phi = 0$. 
The dashed line in the main panel corresponds to $S \sim (U/\kappa)^{1.7}$; the dot-dashed line corresponds to $S = 2 \ln{2} \simeq 1.38$.  
The final time of the time-evolution is $T_{\rm f}\,\kappa \simeq 40.0$ in the main panel and $T_{\rm f}\,\kappa \simeq 2.1$, $2.9$, $6.4$, $15.0$ in the inset, from the bottom to the top curve.
The arrow marks the critical point $U_{\rm c}$. }
\label{fig:entropy}
\end{center}
\end{figure}

We use now the numerical integration of the EOM~(\ref{eq:motiontrace}) to compute the asymptotic value of the order parameter as the interaction strength is increased.
Since we investigate the properties of the homogeneous steady state, it is sufficient to consider a single site ($L = 1$).
We verify that at the end of the time evolution the infinity-norm ${\rm max}\{|\partial_{t} \rho_{\ell;\nu\nu'}|~1\leq\nu,\nu'\leq\nu_{\rm max}\}$ converges towards zero.
In panel (a) of Fig.~\ref{fig:depletion} we plot the condensate expectation value as a function of increasing interaction strength for different, equally spaced times, starting the evolution from an initial fully condensed state ($U = 0$).
We show that the asymptotic values of the order parameter interpolate continuously between the expected values $|\psi|^{2} / n = 1$ for $U = 0$ (coherent state) and $\psi = 0$ (thermal state).
Similarly, the asymptotic values of the entropy $S_{\ell} = - {\rm Tr}[\rho_{\ell} \ln \rho_{\ell}]$ shown in Fig.~\ref{fig:entropy} interpolate between $S = 0$ for the pure state at $U = 0$ and the value $S = 2 \ln 2$ that corresponds to a thermal state with unitary average density.
Between the two extreme values, both quantities present a non-analyticity that develops in time, as the system converges better and better to the steady state.
This effect is clearly a qualitative modification over the smooth crossover that appears in the finite systems of Fig.~\ref{fig:exactcorr} and \ref{fig:exactentropy}.
We may conclude that, in the thermodynamic limit described by the Gutzwiller ansatz, the competition of phase-enhancing dissipation and dephasing due to the interaction leads to non-analytic behavior, i.e.~a nonequilibrium phase transition.

\begin{figure}[t]
\begin{center}
\includegraphics[width=1.00\linewidth]{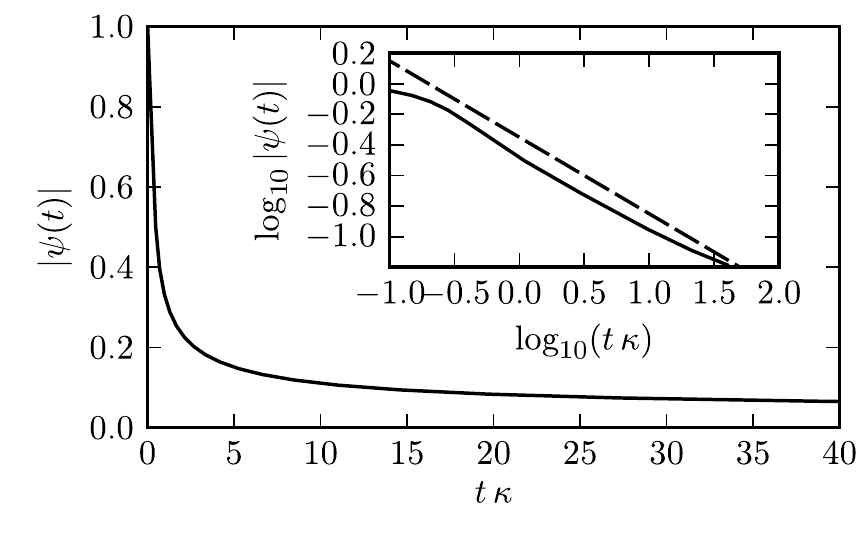}
\caption{Time-evolution of the absolute value of the order parameter in a homogeneous system ($L=1$) for $J = 0.0$, $U/\kappa \simeq 7.2$, $n=1.0$ in linear (main panel) and logarithmic (inset) scale.
The dashed line shows a power-law decay $|\psi(t)| \propto 1/\sqrt{t}$. }
\label{fig:critical}
\end{center}
\end{figure}

To reinforce the previous considerations, we analyze the convergence of the system to the steady state in the proximity of the non-analyticity.
We verify in Fig.~\ref{fig:critical} that, in this case, the convergence to zero takes place with a power law $|\psi(t)| \sim t^{-1 / 2}$, in contrast to the exponential convergence generically granted by the dissipative dynamics.
To be more precise, in the Gutzwiller approach spatial correlations are neglected by construction and we correspondingly observe exponentially decaying correlations, except for the critical point \cite{footnote1}.
To introduce a relation between the time-evolution and the criticality of the system, we note that time sets an energy scale $1 / t$ for the system, which may be viewed as an irrelevant coupling, corresponding to an attractive direction of the fixed point of the system tuned to criticality by choice of the relevant coupling $U / \kappa$.
Indeed, we observe that the non-analyticity corresponding to the critical point develops only in the limit where this scale is removed.
In the vicinity of the critical point the condensate amplitude scales as $|\psi(t)| \sim \exp{(-\lambda_{\rm max} t)} / t^{\alpha}$, where $\lambda_{\rm max}(U / \kappa, J / \kappa)$ is the real part of the largest eigenvalue of the system, vanishing at criticality.
There, the system shows polynomial behavior with universal critical exponent, defined as
\begin{equation}
\alpha = \lim_{t \to \infty} \frac{\partial \log \psi}{\partial \log 1/t} = - \lim_{t \to \infty} \frac{\partial \log \psi}{\partial \log t}~,
\end{equation}
which evaluates to $\alpha = 1/2$ as deduced from Fig.~\ref{fig:critical}, independently of $J/\kappa$ or the mean filling $n$. 
This is the behavior expected within mean field theory and it does not depend on the dimensionality of the system.
While the mean field critical behavior is another clear signature of a second order phase transition taking place in our nonequilibrium system, it is not sufficient to characterize the true universality class of the system. For this purpose, the long wavelength spatial fluctuations would have to be included in an appropriate way, which is not possible within a site decoupling Gutzwiller approach. 
\begin{figure}[t]
\begin{center}
\includegraphics[width=1.00\linewidth]{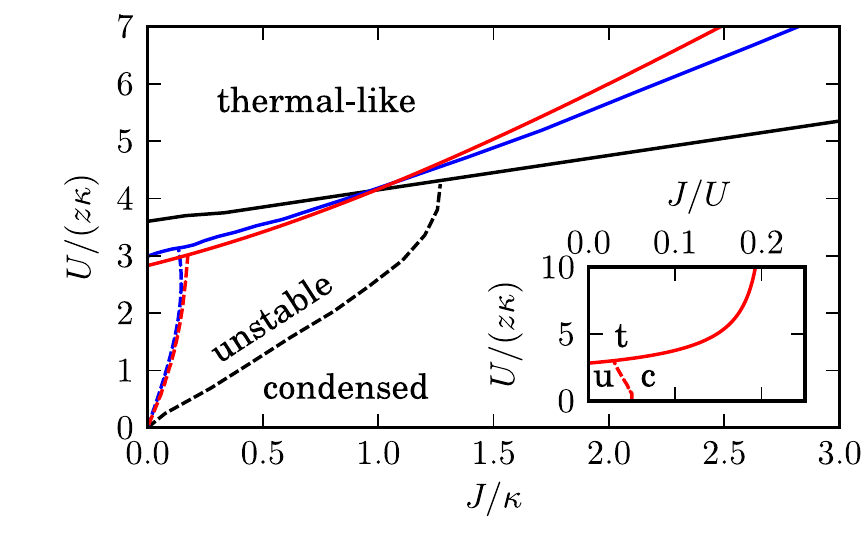}
\caption{Nonequilibrium phase diagram, featuring thermal-like and condensed homogeneous steady states and a domain where the condensed state is unstable.
The black (blue) line is the results of the numerical techniques explained in App.~\ref{app:linthermal} and App.~\ref{app:lingeneral} for $n=1.0$ ($n=0.1$).
The red line is the analytical result in the limit $n \ll 1$ and is repeated in the inset with a different choice of the coordinates for the axes. }
\label{fig:phasediagram}
\end{center}
\end{figure}

Finally, we provide in Fig.~\ref{fig:phasediagram} the position of the critical point as the hopping amplitude $J$ and the interaction strength $U$ are varied, for some values of the density $n$.
To draw the phase border with good accuracy, it is rather awkward to integrate Eq.~(\ref{eq:motiontrace}) repeatedly for long times. 
We resort instead to the analysis of the EOMs linearized around the homogeneous thermal state.
The key observations here is that the thermal-like state (\ref{eq:thrmstate}), contrary to the coherent state (\ref{eq:cohestate}), is always an exact steady state of Eq.~(\ref{eq:motiontrace}).
Its stability property, of course, are different in the two phases and this must appear in the spectrum of the linearized equations.
In the thermal phase, the thermal state is an attractive fixed point, while it is at least a saddle point in the condensed state. 
Around the stable fixed point our approach analyzes the linear response of the system to small perturbations, while around a generic fixed point it assesses the existence of unstable directions in the phase space.
In generic nonlinear equations, two or more (meta-)stable steady states could occur. 
This does not seem to be the case in the present model based on our numerical findings, and is also reasonable given the fact that the evolution of the density matrix elements $\partial_t \rho_{\ell;\nu\nu'}$ is generated by a quadratic form. 
(One could also conceive a steady state that is unstable, but still is stable at any order in the Taylor series: this again does not happen here.)
The derivation of the linearized EOMs is deferred to App.~\ref{app:linthermal}, but we mention here that this method allows us to test the stability of several hundred pixels in the phase diagram, in a short computational time, truncating the Hilbert space to a maximum number of particles $\nu_{\max}$ larger than $10^{2}$.
The latter condition is particularly important because the diagonal elements of the thermal state decrease only as a power law with the Fock index $\nu$.

Quite nicely, this approach allows us also to obtain the analytical form of the phase boundary in the limit $n \ll 1$ of small particle filling.
To this end, we consider the linearized equations (\ref{eq:instabEq}) around the thermal state with onsite Fock spaces  truncated at occupation number $\nu_{\rm max} = 2$, appropriate for sufficiently low filling but still capable of describing the effect of interactions.
Then we have a $3 \times 3$ linear system where one eigenvalue is $i U / 2 \kappa$ and corresponds to the $\nu = 0$ state.
The other two eigenvalues $\lambda_{-}$, $\lambda_{+}$ can be computed straightforwardly.
Keeping only the first order in the average density, one eigenvalue $\lambda_{-}$ has negative real part while for the other ${\rm Re}[\lambda_{+}] = 4 n \kappa (4 J^{2} - U^{2} + 4 J U + 32 \kappa^{2}) / [(2 J + U)^{2} + 64 \kappa^{2}]$.
The curve in the $(J,U)$ plane where $\lambda_{+} = 0$ defines the phase border, i.e.~where the thermal state changes from stable to unstable.
In the current approximation it reads
\begin{equation}\label{eq:phaseborder}
U = 2 J + \sqrt{8 J^{2} + U_{\rm c}^{2}}~,
\end{equation}
with $U_{\rm c} = \sqrt{32} \kappa$ the critical point at $J = 0$ ($z = 1$).
This expression is shown as a red solid line in the main panel and in the inset of Fig.~\ref{fig:phasediagram}. 
Below in Sec.~\ref{sec:smalldens} we point out an alternative way of deriving this result. 
By reducing the analysis of the equations of motion to the first three Fock states it is also possible to extrapolate the position of the critical point for variable density, as shown in the panel (c) of Fig.~\ref{fig:depletion}.

Two features of the phase diagram of Fig.~\ref{fig:phasediagram} are dramatic manifestations of the system being out of equilibrium.
First, even in the limit of vanishing dissipation the system presents a transition from a thermal-like to a condensed steady state at $J / U = (\sqrt{2}  -1)/2 \simeq 0.21$.
This picture does not connect to the phase transition between the Mott state and the superfluid state that characterizes the phase diagram of the equilibrium Bose-Hubbard system.
Formally, such difference is possible because the limits $\kappa \to 0$ and $t \to \infty$ do not commute and defeats the naive expectation that a small dissipation applied to a system does not wash out entirely its zero-temperature phase diagram.
A more physical explanation is the fact that the engineered dissipation in the Liouvillian~(\ref{eq:liouvillian}) does not respect the principle of detailed balance, as discussed above, and hence its action is radically different from a bath that induces thermalization.
A notable example of the latter case is a Josephson junction array coupled to a dissipative bath \cite{JJarrays}, which suitably chosen can stabilize the superconducting ordered phase.
The profound difference with respect to our setup stems from the fact that there the system plus bath is in global thermodynamic equilibrium and the associated phase transitions (thermal or quantum) are equilibrium ones.

Second, we note the absence of commensurability effects.
At the mean field level and at zero temperature, one expects that any mechanism suppressing superfluidity leads to a Mott state for commensurate filling.
In fact, given the density matrix $\rho_{\ell} = \sum_{\nu} |\nu\rangle_{\ell}\langle \nu|~p_{\nu} $ without off-diagonal order (which represents long-range order in the mean field approximation), the only solution for the steady state is a Mott state $p_{\nu} = \delta_{\nu,n_0}$ with quantized particle number $n_{0}$, because the conditions $\sum_{\nu} p_{\nu} =1$ (normalization)  and $\sum_{\nu} p_{\nu}^2 = 1$ (purity at zero temperature) must hold simultaneously on each site of the lattice.
In contrast, in our case the dephasing of correlations induced by the interaction (see discussion in Sec.~\ref{ssec:competition}) leads to a diagonal steady state that does not obey an additional purity constraint. 
The dephasing of the coherent initial state makes the density matrix more and more diagonal, but does not lead to an additional localization in Fock space \cite{footnote2}.
We also observe directly that the phase diagram in Fig.~\ref{fig:phasediagram} does not change qualitatively from $n < 1.0$ to $n = 1.0$.
In conclusion, the nonequilibrium quantum phase transition presented here shares features of equilibrium quantum phase transitions in that it is interaction driven, and of classical phase transitions in that the ordered phase terminates in a thermal state. 

\subsection{The role of the chemical potential}
\label{ssec:chempot}

In our formulation of the many-body nonequilibrium problem, we make use of a chemical potential, which at first sight is a concept tightly bound to thermodynamic equilibrium. 
In order to better understand its meaning and practical use in the present setting,  we would like to compare our situation to the thermodynamic equilibrium of an interacting Bose gas, say in the continuum, in which the temperature is low enough that spontaneous symmetry breaking takes place. 
Here, the role of the chemical potential is twofold: first, it conceptually fixes the (average) particle number; second, its actual value is chosen to fulfill an equilibrium condition, namely to remove homogeneous terms which are linear in the fluctuation operators which occur upon expansion about a condensed ground state, a condition which in turn is equivalent to the existence of a gapless Goldstone mode.

In our nonequilibrium case, there is no coincidence of these two roles of the chemical potential; its choice is only connected to the second one. 
Indeed, the average particle number is an exactly conserved quantity of our mean field master equation. 
Hence, the average particle number is fixed via the initial state. 
However, the introduction of a chemical potential term is needed to ensure the existence of a steady state, in which all time-local correlation functions become time independent. 
Mathematically, this choice can be made by requiring that the equation of motion for the spatially homogeneous fluctuation $\psi_{\ell}(t) - \psi(0)$ experiences no ``driving'' for all $\ell$, i.e.~there is no constant term generating its evolution. 
This condition on the equation of motion is precisely analogous to the condition of the vanishing of the coefficient of the term linear in the fluctuation in a Hamiltonian. 

\subsection{Dynamical instability in the nonequilibrium phase diagram}
\label{ssec:stability}
\begin{figure}[t]
\begin{center}
\includegraphics[width=1.00\linewidth]{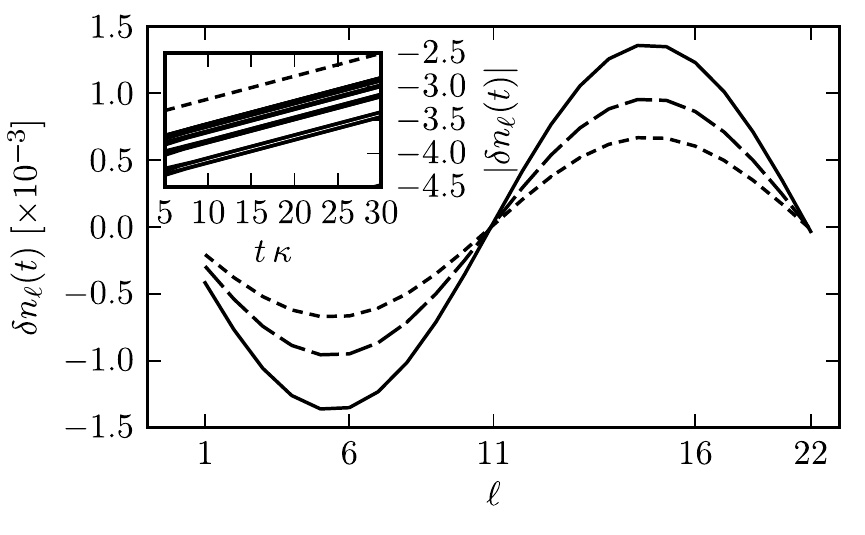}
\caption{Time-evolution of the density wave generated by the dynamical instability, according to the mean field EOM (\ref{eq:motiontrace}), for $L=22$, $n=1.0$, $J = 0.0$, and $U / \kappa = 3.0$.
The main panel shows the fluctuations $\delta n_{\ell}$ of the local density $n_{\ell}$ with respect to the average value (short dash, long dash, solid, from the earlier to the later time).
The inset shows the absolute values of the fluctuations in logarithmic scale (solid) and the prediction of the growth rate obtained with the numerical linearization of the equations of motion around the steady state (dashed).}
\label{fig:densitywave}
\end{center}
\end{figure}

Having classified the homogeneous steady states of the system and presented the phase border in Fig.~\ref{fig:phasediagram}, we consider now the stability of the condensed phase with respect to inhomogeneous perturbations.
Although in general we expect that an initial inhomogeneous system is driven towards the homogeneous dark state defined by the Liouvillian (as discussed in Sec.~\ref{sec:themodel} and shown for a paradigmatic case in Fig.~\ref{fig:condensation3d}) we cannot discard the possibility that the competition between unitary and dissipative dynamics destabilizes the convergence process in some range of parameters.
The stability of the homogeneous steady state can be tested numerically by subjecting the steady state to a small perturbation and monitoring the dynamics following the perturbation.
Due to the nonlinearity of the dynamics described by Eq.~(\ref{eq:motiontrace}), the results of this analysis depend on the kind of perturbation that is applied to the system.
From the theory of the linear response of many-body systems, we know that the spectrum of the collective excitations is obtained by studying the density-density response function, hence we perturb the density matrix of the system with the addition of a coherent component $\delta \rho_{\ell}$ that oscillates periodically in space $\propto \cos{(2\pi \ell / L)}$ with the maximum wavelength allowed by the size of the lattice.
In Fig.~(\ref{fig:densitywave}) we show that, for specific choices of the parameters, the magnitude of the perturbation $\delta n_{\ell}$ is not damped, but increases exponentially in time as $e^{\gamma t}$, with $\gamma > 0$, preserving its shape.
The robustness of the spatial profile $\delta n(t)$ and the fact that it increases in time with a well-defined instability exponent $\gamma$ classifies this phenomenon as an unstable mode.
The system then manifests a dynamical instability, characterized by the wavelength of the charge density wave show in Fig.~\ref{fig:densitywave}.
A more thorough analytical understanding of the origin of this instability is deferred to Sec.~\ref{sec:instability}, after a suitable approximation scheme is developed.

To delimitate systematically the range of parameters where the dynamical instability takes place, we linearize the EOM (\ref{eq:motiontrace}) around the steady state and study the spectrum of eigenvalues, analogously to the procedure performed in the previous section to find the border of stability for the thermal state.
In this case, however, the substantial analytical simplifications deduced in the previous case cannot be applied because the form of the condensed steady state is not known in general and, in any case, the EOMs for the entries of the density matrix have a much more intricate structure of couplings.
Hence we resort to the numerical procedure outlined in App.~\ref{app:lingeneral}, which provides the shape of the unstable domains shown in Fig.~\ref{fig:phasediagram} with dashed lines.
We see that an unstable domain exists, that includes the whole $J = 0$ axis where the thermal state is not stable, and is larger as the interaction strength increases.
The phase transition from the thermal to the condensed state is then substituted by a transition from the thermal to the unstable phase for small enough values of $J$.
The border between the condensed and the unstable regime in the proximity of the origin is characterized by a fixed ratio $J / U$, as can be clearly seen in the inset of Fig.~\ref{fig:phasediagram}.
To interpret this fact, we note that the condensed eigenstate of the Liouvillian (\ref{eq:liouvtrace}) is also an eigenstate of the kinetic term in Eq.~(\ref{eq:localhamiltonian}), so $J$ is the strength of an unitary term ``compatible'' with the engineered dissipation. 
On the contrary, interaction $U$ and dissipation compete strongly, as discussed in detail above.
We may argue then that, in general, the steady state of the Liouvillian can be made more robust by increasing the compatible energy scales with respect to the competing energy scales in the Hamiltonian.

\section{Asymptotic behavior and criticality in the limit of small density}
\label{sec:smalldens}
\begin{figure}[t]
\begin{center}
\includegraphics[width=1.00\linewidth]{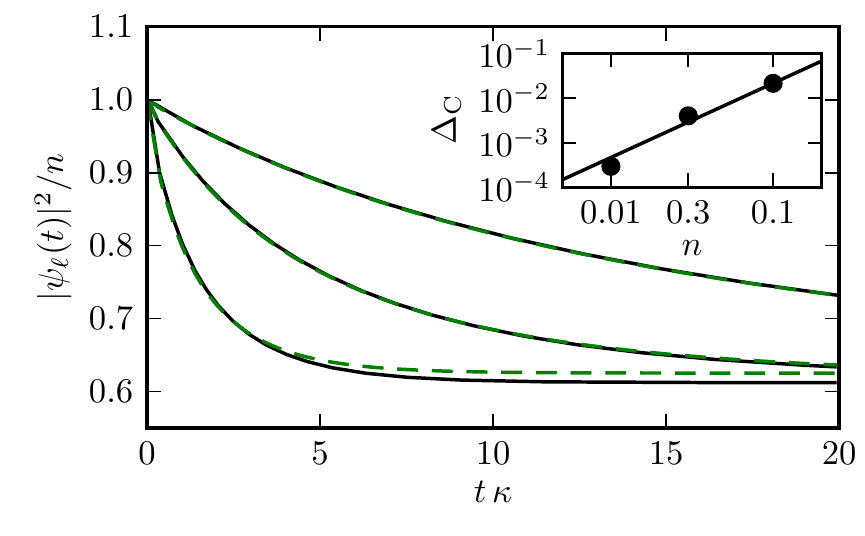}
\caption{ The main panel compares the time-evolution of the condensate fraction according to the mean field EOM (\ref{eq:motiontrace}) and the EOM for small density (solid and dashed lines, respectively) for $n=0.1$, $0.03$, and $0.01$ from bottom to top in a homogeneous system ($L = 1$).
The inset shows in a bilogarithmic scale the difference $\Delta_{\rm C}$ between the results at the final time $T_{\rm f} = 20.0$ for the three values of the density.}
\label{fig:reduced}
\end{center}
\end{figure}

We present now several analytical arguments that confirm the nonequilibrium phase diagram presented in Fig.~\ref{fig:phasediagram}.
We consider the equations of motion for the most general normal ordered correlation functions of our system $c_{p,q,\ell} = \langle \hat{b}_{\ell}^{\dag \, p} \hat{b}_{\ell}^{q} \rangle$.
In general, these correlation functions form an infinite nonlinear coupled hierarchy. 
However, we find that the \emph{a priori} infinite hierarchy of mean field equations for the onsite correlation functions exhibits a finite subset which decouples from the rest in the limit $n \to 0$. 
Importantly, we find that all qualitative effects of the nonequilibrium phase diagram are contained in this reduced system of equations of motion, making this limit an important tool for an analytical understanding of the physics of our system.  
In Fig.~\ref{fig:reduced} we show that the results obtained with the numerical integration of the reduced equations compares quantitatively very well with the full equations for average densities even larger than $n = 0.1$, i.e. a realistic figure for typical fillings in optical lattices.

For the full system of correlation functions, the general form for the equation of motion $\partial_{t} c_{p,q,\ell}$ in closed form is reported in Eq.~(\ref{eq:motionanycorr}). 
We observe that $c_{p,q,\ell}$ couples to the correlation functions $c_{p+1,1,\ell}$ and $c_{p,q+1,\ell}$, thereby not allowing for a truncation to small subset of correlation functions \emph{a priori}. 
However, progress can be made by introducing a power counting $\hat{b}_{\ell} \sim n^{1/2}$. 
This gives an upper bound on the true powers of the operators, as correlation functions in the state without phase symmetry breaking with $p \neq q$ vanish exactly. 
The leading power for the time evolution of a correlation function is consequently given by $c_{p,q,\ell} \sim n^{(p+q)/2}$. 
Writing down the system of equations of motions (\ref{eq:corrset}), it is then obvious that this system closes if we truncate the equations of motion for $\langle \hat{b}_{\ell}^{2} \rangle$, $\langle \hat{n}_{\ell} \hat{b}_{\ell} \rangle$ each of the equations at their leading power, thereby physically corresponding to a low density expansion. 

We use this result first to obtain analytically the critical exponent $\gamma$ discussed above.
For a homogeneous system with $J=0$ the equations of motion read
\begin{eqnarray}\label{eq:corrhom}
\partial_{t} \langle \hat{b}_{\ell} \rangle & = & inU \langle \hat{b}_{\ell} \rangle + (-iU + 4\kappa) \langle \hat{n}_{\ell} \hat{b}_{\ell} \rangle - 4 \kappa \langle \hat{b}_{\ell} \rangle^{\ast} \langle \hat{b}_{\ell}^{2} \rangle  \nonumber \\
\partial_{t} \langle \hat{n}_{\ell} \hat{b}_{\ell} \rangle & = & 8 n \kappa \langle \hat{b}_{\ell} \rangle + (-iU + i n U - 8 \kappa) \langle \hat{n}_{\ell} \hat{b}_{\ell} \rangle \nonumber \\
\partial_{t} \langle \hat{b}_{\ell}^{2} \rangle & = & (-iU + 2 i n U - 8 \kappa) \langle \hat{b}_{\ell}^{2} \rangle + 8 \kappa \langle \hat{b}_{\ell} \rangle^{2}~,
\end{eqnarray}
with the chemical potential $\mu = n U$ resulting from the steady state condition, cf.~the discussion following Eq.~(\ref{eq:dissgpe}) \cite{footnote3}.
The exponential convergence of the system to its steady state is described by the linear contribution, which corresponds to a ``mass'' or ``gap'' in the jargon of statistical mechanics.
We are interested in the critical regime of the dynamics, in which the exponential convergence is suppressed due to the vanishing of the gap term and the system is driven to its fixed point by the residual nonlinear contributions to the equations of motion.
We focus on the linear equations for $\langle \hat{b}_{\ell} \rangle$ and $\langle \hat{n}_{\ell} \hat{b}_{\ell} \rangle$, with eigenvalues $\lambda_{1}$, $\lambda_{2}$ and we consider $\langle \hat{b}_{\ell}^{2} \rangle$ as a nonlinear driving term.
In the critical regime, in which $\lambda_{1} = 0$, the secular equation for the eigenvalue fixes the critical value of the interaction strength $U^{2} = 32 \kappa^{2} / (1-n)$ [that coincides with $U_{\rm c}$ defined after Eq.~(\ref{eq:phaseborder}) for $n \ll 1$].
The eigenvector corresponding to $\lambda_{1}$ is $(i U - 4 \kappa, i n U)$ and hence the EOM for the quantity $\varphi_{1} = (i U - 4 \kappa) \langle \hat{b}_{\ell} \rangle + i n U \langle \hat{n}_{\ell} \hat{b}_{\ell} \rangle$ does not feature linear terms and reads $\partial_{t} \varphi_{1} = -4 \kappa (i U - 4 \kappa) \langle \hat{b}_{\ell} \rangle^{\ast} \langle \hat{b}_{\ell}^{2} \rangle$.
A similar quantity $\varphi_{2}$ built with the eigenvector corresponding to $\lambda_{2}$, instead, features a regular exponential decay in time.
The time-evolution of $\langle \hat{b}_{\ell} \rangle$ is given in general by a linear combination $\varphi_{1}$ and $\varphi_{2}$ but, for long times, the exponentially decaying contribution of the latter is negligible and hence we can substitute $\langle \hat{b}_{\ell} \rangle \sim \varphi_{1} / (i U - 4 \kappa)$ into the EOM for $\varphi_{1}$.
At this point we make an adiabatic approximation by solving the equation $\partial_{t} \langle \hat{b}_{\ell}^{2} \rangle = 0$ for $\langle \hat{b}_{\ell}^{2} \rangle$ and substituting the result into the EOM for $\varphi_{1}$, that yields $\partial_{t} |\varphi_{1}|^{2} = - |\varphi_{1}|^{4} / (3 \kappa)$.
From the solution $|\varphi_{1}|^{2} = 3 \kappa / t$, using the expressions above, we finally demonstrate the critical behavior $|\langle \hat{b}_{\ell} \rangle| = 1 / (4 \sqrt{t \kappa})$.

We focus now on the asymptotic behavior of the system to demonstrate the depletion of the condensate fraction shown in Fig.~\ref{fig:depletion}.
To this end, it is more convenient to introduce the fluctuation operators $\delta \hat{b}_{\ell} \equiv \hat{b}_{\ell} - \psi_{\infty}$ where $\psi_{\infty}$ is the (unknown) order parameter in the homogeneous steady state.
The EOMs for the ``connected'' correlation functions built with the fluctuation operator (see App.~\ref{app:eqmotion}) decouple naturally into linear equations that involve the connected parts of the mean fields (\ref{eq:corrset}), the density fluctuation $\langle \delta \hat{b}_{\ell}^{\dag} \delta \hat{b}_{\ell} \rangle$, and $\psi_{\infty}$ as a parameter.
In the homogeneous steady state, in which $\langle \delta \hat{b}_{\ell} \rangle = 0$ (see App.~\ref{app:eqmotion}) we have
\begin{widetext}
\begin{eqnarray}\label{eq:corrsteadystate}
\partial_{t} \langle \delta \hat{b}_{\ell} \rangle & = & 
 -i U \psi_{\infty} {\langle \delta \hat{b}_{\ell}^{2} \rangle}-2 (i U -  4 \kappa) \psi_{\infty} {\langle \delta \hat{b}_{\ell}^{\dag} \delta \hat{b}_{\ell} \rangle}-(i U - 4 \kappa) {\langle \delta \hat{b}_{\ell}^{\dag} \delta \hat{b}_{\ell}^{2} \rangle} -i U \psi_{\infty}^3+i {\mu_{1}} \psi_{\infty} = 0 \nonumber \\
\partial_{t} \langle \delta \hat{b}_{\ell}^{\dag} \delta \hat{b}_{\ell} \rangle & = & 
 i U \psi_{\infty}^2 {\langle \delta \hat{b}_{\ell}^{2} \rangle}-i U \psi_{\infty}^2 {\langle \delta \hat{b}_{\ell}^{2} \rangle^{\ast}}+(i U - 4 \kappa) \psi_{\infty} {\langle \delta \hat{b}_{\ell}^{\dag} \delta \hat{b}_{\ell}^{2} \rangle}-(i U + 4 \kappa) \psi_{\infty} {\langle \delta \hat{b}_{\ell}^{\dag} \delta \hat{b}_{\ell}^{2} \rangle^{\ast}}-16 \kappa  \psi_{\infty}^2 {\langle \delta \hat{b}_{\ell}^{\dag} \delta \hat{b}_{\ell} \rangle}  = 0 \nonumber \\
\partial_{t} \langle \delta \hat{b}_{\ell}^{2} \rangle & = & 
- (i U + 8 \kappa + 4 i J) {\langle \delta \hat{b}_{\ell}^{2} \rangle}-i U \psi_{\infty}^2  = 0 \nonumber \\
\partial_{t} \langle \delta \hat{b}_{\ell}^{\dag} \delta \hat{b}_{\ell}^{2} \rangle & = &
- (i U + 8 \kappa + 2 i J) {\langle \delta \hat{b}_{\ell}^{\dag} \delta \hat{b}_{\ell}^{2} \rangle}-2 (i U + 4 \kappa) \psi_{\infty} {\langle \delta \hat{b}_{\ell}^{\dag} \delta \hat{b}_{\ell} \rangle^{\ast}}  = 0~.
\end{eqnarray}
\end{widetext}
The nonlinearity of the theory is now entirely contained in the identity $n = \langle \delta \hat{b}_{\ell}^{\dag} \delta \hat{b}_{\ell} \rangle \vert_{\psi_{\infty}} - |\psi_{\infty}|^{2}$, whence the value of the order parameter is obtained once the other equations have been solved.
Practically, in the solution we can assume that the order parameter is real without loss of generality and fix the chemical potential to ensure that the fluctuation of the density has vanishing imaginary part.
The full result for the depletion then reads
\begin{equation}\label{eq:analytdepletion}
\frac{\psi_{\infty}^{2}}{n} = 1 - \frac{U^2 (U^2 + 4 J U + 4 J^2 + 64 \kappa^2)}{8[U^2 c_{2} + U c_{1} + 8 J^4 + 96 J^2 \kappa^2 + 128 \kappa^4]}~,
\end{equation}
with $c_{2} = 4 \kappa^2 + 3 J^2$ and $c_{1} = 12 J^3 + 64 J \kappa^2$.
For $J=0$ ($z = 1$) this reduces to the simple inverted parabolic profile
\begin{equation}
\left ( \frac{\psi_{\infty}^{2}}{n} \right )_{J = 0}  = 1 - \left(\frac{U}{U_c}\right)^2~,\quad U_c = 4 \sqrt 2 \kappa.
\end{equation}
These analytical results are compared to the numerical solution of the equations of motion in the panel (b) of Fig.~\ref{fig:depletion}, demonstrating an excellent agreement between the two computations.
The quadratic behavior of the depletion may be contrasted with the thermal depletion in a free equilibrium Bose-Einstein condensate in three dimensions, where $|\psi|^2/n = 1 - (T/T_c)^{3/2}$ with $T_c = 2\pi/M(n/\zeta(3/2))^{2/3}$ with $M$ the particle's mass and $\zeta$ the Riemann zeta function. 
The quadratic behavior is reminiscent of a harmonically trapped Bose-Einstein in two dimensions.
It is noteworthy that the phase border that one obtains by requiring $\psi_{\infty} = 0$ in Eq.~(\ref{eq:analytdepletion}) coincides exactly with the small-density result in Eq.~(\ref{eq:phaseborder}) in the linear response strategy of Sec.~\ref{ssec:phases}.

\section{Fluctuations-induced dynamical instability}
\label{sec:instability}

\subsection{Linear response}
\label{ssec:linresponse}

We are now in the position of providing a clear-cut explanation of the dynamical instability unveiled in Sec.~\ref{ssec:stability}, transposing at the level of the correlation functions the instability analysis around the steady state used to delineate the unstable domains in Fig.~\ref{fig:phasediagram}.
The starting point are the EOMs for the connected correlation functions outlined in App.~\ref{app:eqmotion}, and the knowledge of the homogeneous steady state $\rho_{\infty}$ demonstrated in the previous section.
Then we expand the density matrix of the system as $\rho_{\ell}(t) = \rho_{\infty} + \Delta \rho_{\ell}(t)$ and consider the fluctuations of the connected correlation functions
\begin{eqnarray}
\langle \delta \hat{b}_{\ell}^{\dag p} \delta \hat{b}_{\ell}^{q} \rangle & = & \langle \delta \hat{b}_{\ell}^{\dag p} \delta \hat{b}_{\ell}^{q} \rangle_{\infty} +  \Delta \langle \delta \hat{b}_{\ell}^{\dag p} \delta \hat{b}_{\ell}^{q} \rangle \nonumber \\
& = & {\rm Tr}[ \delta \hat{b}_{\ell}^{\dag p} \delta \hat{b}_{\ell}^{q} \rho_{\infty}] +  {\rm Tr}[ \delta \hat{b}_{\ell}^{\dag p} \delta \hat{b}_{\ell}^{q} \Delta \rho_{\ell}(t) ]~. 
\end{eqnarray}
The EOMs are then linearized in the fluctuations $\Delta \langle \delta \hat{b}_{\ell}^{\dag p} \delta \hat{b}_{\ell}^{q} \rangle $ and the values of $\langle \delta \hat{b}_{\ell}^{\dag p} \delta \hat{b}_{\ell}^{q} \rangle_{\infty}$, given by Eq.~(\ref{eq:corrsteadystate}), are such that all the driving (i.e.~constant) terms vanish upon the correct choice of the chemical potential outlined above.
We remark that the fluctuation $\Delta \langle \delta \hat{b}_{\ell} \rangle $ does not vanish as the system is not in its steady state (see App.~\ref{app:eqmotion}) and that local fluctuations $\Delta \langle \delta \hat{b}_{\ell}^{\dag} \delta \hat{b}_{\ell} \rangle$ of the density modify the flat distribution $\langle \hat{n}_{\ell} \rangle \equiv n$ of the steady state.
Then we take the Fourier transform of the EOMs and rewrite the linear system in terms of the connected correlation functions in momentum space $\Delta \langle \delta \hat{b}^{\dag p} \delta \hat{b}^{q} \rangle_{q} \propto \sum_{\ell} e^{i q \ell} \langle \delta \hat{b}_{\ell}^{\dag p} \delta \hat{b}_{\ell}^{q} \rangle$.
It is important to notice that the Fourier transform of the correlation functions is not simply related to the correlation functions of the operators in momentum space, except for the first order correlation where it holds $\Delta \langle \delta \hat{b} \rangle_{q} = \Delta \langle \hat{b}_{q} \rangle$ and $\Delta \langle \delta \hat{b} \rangle_{q}^{\ast} = \Delta \langle \hat{b}_{-q} \rangle$.
(We may denote such correlations $\Delta \psi_{q}$ and $\Delta \psi_{-q}$, respectively, because the fluctuations of the order parameter vanish on the steady state by construction.)
Since the instability shown in Fig.~\ref{fig:densitywave} takes place at small momenta, in performing the Fourier transform we focus on the central region of the Brillouin zone and substitute the occurrences of the discrete Laplacian $\Delta_{\ell}u \equiv u_{\ell+1} - 2 u_{\ell} + u_{\ell - 1}$ with the parabolic dispersion $-q^{2} u_{q}$.
\begin{figure}[t]
\begin{center}
\includegraphics[width=1.00\linewidth]{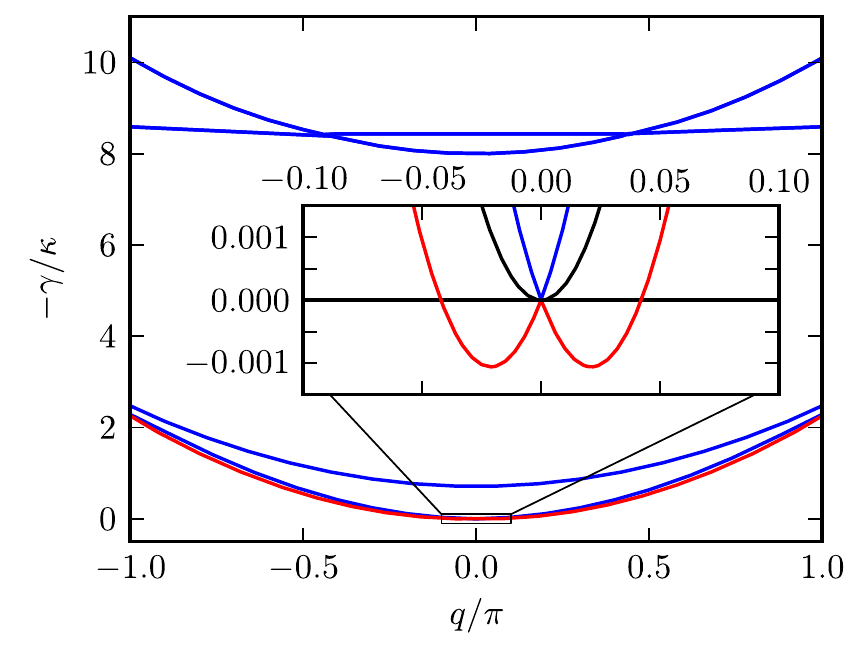}
\caption{Real part of the eigenvalues $\lambda = -i \omega + \gamma$ of the linearized EOM for $J=0$, $n=0.1$, and $U = 1.0 \kappa$. }
\label{fig:spectrum}
\end{center}
\end{figure}

The linear system of EOMs takes the form of a $7 \times 7$ matrix (the $3$ complex correlation functions, their complex conjugates, and the real density fluctuation) whose eigenvalues $\lambda = -i\omega + \gamma$ give the $q$-dependent spectrum of the system.
The eigenvectors of the system correspond to modes that evolve as $\delta \rho_{\ell}(t) = \delta \rho_{\ell}(0) e^{-i \omega t} e^{+ \gamma t}$, which are stable (unstable) if $\gamma < 0$ ($\gamma > 0$).
The real part of the spectrum for a typical choices of parameters within the unstable domain is shown in Fig.~\ref{fig:spectrum}.
The spectrum features: i) two doubly-degenerate strongly decaying modes ($\gamma/\kappa \simeq 9.0$) that project mainly on the third-order correlation functions; ii) one decaying mode ($\gamma / \kappa \simeq 1.5$) that projects mainly on the density fluctuation; iii) two low-lying modes generated by an admixture of the first-order correlation functions.
The latter modes are magnified in the inset of Fig.~\ref{fig:spectrum}.
The lower mode gives $\gamma > 0$ in a small interval around $q = 0$, hence proving the existence of unstable modes with well-defined momentum.
The domain where an unstable mode exists in this approximation is delimited in Fig.~\ref{fig:phasediagram} by a red dashed line.

In general, the decay rate of the modes i) and ii) is proportional to ${\cal O}(\kappa)$ and ${\cal O}(\kappa n)$, respectively, as it appears from inspection of the linearized EOMs.
The clear separation of the dissipative part of the spectrum into groups of modes that have largely different decay rate at low momenta suggests that an adiabatic elimination of the fastest modes can be performed to bring the $7 \times 7$ linear system in a more compact form.
In this way we obtain a renormalized spectrum of the weakly dissipative single particle modes, where the instability is encoded.
In general, the adiabatic elimination in a system
\begin{equation}
\partial_{t} u_{\rm F} = F[u_{\rm F}, u_{\rm S}], \quad 
\partial_{t} u_{\rm S} = G[u_{\rm F}, u_{\rm S}]~,
\end{equation}
with fast ($u_{\rm F}$) and slow ($u_{\rm S}$) modes consists of solving $F[u_{\rm F}, u_{\rm S}] = 0$ for $u_{\rm F}$ and using the result into the second equation that becomes $\partial_{t} u_{\rm S} = G[F^{-1}[0, u_{\rm S}], u_{\rm S}]$.
The application of the procedure introduces new terms $F^{-1}[0, u_{\rm S}]$ that renormalize the equation of the slow modes.
We apply the procedure once to eliminate the $\gamma \sim {\cal O}(\kappa)$ modes and then again to eliminate the $\gamma \sim {\cal O}(\kappa n)$ modes.
Since the border of the unstable domain extends to the origin of the phase diagram in $J$, $U$ (see Fig.~\ref{fig:phasediagram}), to understand the phenomenon underlying the instability it is enough to perform the algebraic manipulations to the first order in $J$ and $U$.
We obtain a renormalized $2 \times 2$ linear system for the time derivative $\partial_{t} ( \Delta \psi_{q} , \Delta \psi_{-q}^{\ast})$ of the fluctuations of the order parameter in time, which reads
\begin{equation}
\left ( \!\!\! \begin{array}{cc}
- i (n U + \varepsilon_{q}) - \kappa_{q} + r_{q} &
- i n U + s_{q} \\
+i n U + s^{\ast}_{q} &
+i (n U + \varepsilon_{q}) - \kappa_{q} + r^{\ast}_{q}
\end{array} \!\!\! \right )~.
\end{equation}
Here, $\kappa_{q} = 2\kappa q^{2} (2 n + 1)$ is the ``bare'' quadratic decay rate that follows from the analysis of the linear correlations only, for small interaction and nonzero hopping amplitude, and is shown as a black solid line in the inset of Fig.~\ref{fig:spectrum}.
$\varepsilon_{q} = J q^{2}$ is the low momentum kinetic energy.
Finally, $r_{q} = q^{2} (15 n J U / \kappa + 22 i n U) / 32$ and $s_{q} = - q^{2} (n J U / \kappa + 7 i n U) / 16$ are the terms that renormalize the slow modes obtained by the adiabatic elimination.
Without the renormalizing terms the $2 \times 2$ system displays the structure of a Bogoliubov equation for the condensate modes, with diagonal dissipation $\kappa_{q}$.
We point out that a standard quadratic theory can reproduce the Bogoliubov-type EOM but necessarily misses the renormalizing terms that are due to third-order local correlations, and thereby the entire physics of the dynamical instability. 
The latter is thus a clear fluctuation-induced beyond-mean-field effect.
The eigenvalue of the linear system, which approximates the lower mode in the inset of Fig.~\ref{fig:spectrum}, reads explicitly
\begin{equation}
-i \omega_{q} + \gamma_{q} = -i q \sqrt{n U (8 J - 9 n U)} / 2 - \kappa_{q} + 15 q^2 n J U / (32 \kappa)~.
\end{equation}
If $J > 8 n U / 9$ we can identify $c = \sqrt{n U (8 J - 9 n U)} / 2$ as the speed of sound $\omega_{q} = c |q|$ of the dissipatively-created condensate and we also find a modified decay rate for the modes that is quadratic in the momentum.
However, as $J$ increases, the square root becomes imaginary, the contribution of the dispersion to $\omega_{q}$ vanishes and the decay rate of the modes is modified by a non-analytic term $\sim |q|$, which is positive and dominates over the contribution $\sim q^2$  at low momentum.

The resulting physical picture for the dynamical instability in the dissipatively-driven condensate is then  an interplay of short time quantum and long time classical fluctuations.
Purely local, i.e.~temporal (or quantum) fluctuations captured in the Gutzwiller approach renormalize the complex spectrum at short times. 
This prepares the ground for spatial (or classical) fluctuations taking over at later times and responsible for the dynamical instability.
We remark that the Gutzwiller approximation that we adopt is suited to describe both i) the onsite fluctuations that are described without approximation in a fully quantum local Hilbert space; ii) long-wavelength fluctuations of the coherence that receive the main contribution from the disconnected part, which is computed solely in terms of the order parameter.

It is important to realize that the instability arises at weak coupling already, where the system is well described by the inhomogeneous Gutzwiller mean field theory. 
Without referring to any approximation, it is clear that the instability is due to a renormalization of the single particle (complex) excitation spectrum, and thus encoded in the evolution of $( \Delta \psi_{q} , \Delta \psi_{-q}^{\ast})$.
The exact equation of motion for these quantities is a nonlinear equation, with in general nonlocal spatial correlations. 
The Gutzwiller approximation factorizes the correlations functions in real space, but treats onsite correlations exactly. 
The factorization in real space is justified at weak coupling (large condensate) because the dominant scattering processes are those for opposite momenta off the macroscopically occupied condensate. 
However,  the factorization of the onsite correlation functions (as in the GPE discussed in Sec.~\ref{ssec:phases}) would not allow to capture the physics of the dynamical instability.
In summary, because the dynamical instability is a weak coupling phenomenon where our approximations can be rigorously justified, we can rule out the possibility of the effect being an artifact of approximations. 
Of course, the above analysis has the status of a linear response theory, and thus is not able to shed light the fate of the system in the dynamically unstable regime; we will address this question in Sec.~\ref{ssec:nonlineresponse}.

We note now that the exact Liouville operator (\ref{eq:liouvillian}) features the dark state also in the case of a single particle on the lattice, corresponding to the limit $n \to 0$ with $N = 1$ and $L \to \infty$. 
Our approximation yields correctly the existence of the dark state at $q = 0$, since the expression for the bare quadratic decay rate reads $\kappa_{q} = 2 \kappa q^{2}$ in the limit $n \to 0$.
On the contrary, the Bogoliubov approximation used in Ref.~\cite{DMKKBZ2008} gives $\kappa_{q} \simeq 4 n \kappa q^{2}$ and hence, in the limit $n \to 0$, does not yield the correct unique dark state.
The discrepancy between the two methods lies in the fact that, in the Bogoliubov approximation, it holds $\langle \hat{b}_{\ell} \hat{b}_{\ell}^{\dag} \rangle \simeq \langle \hat{b}_{\ell}^{\dag} \hat{b}_{\ell} \rangle$ leading to $2n+1 \simeq 2n$.
The additional occurrence of unity that is neglected in the Bogoliubov approximation is indeed essential to the existence of the dark state and is correctly captured by the Gutzwiller approximation.
Finally, we remark that, in contrast to the dissipative part, the bare theory for the unitary part of the time evolution, where two-point correlations are not included, coincides with the GPE augmented with linear fluctuations. 
This is rooted in the precise nature of the nonlinearities occurring in the dissipative and unitary part.

\subsection{Spontaneous breaking of translational symmetry at large times}
\label{ssec:nonlineresponse}

Because of the dynamical instability demonstrated above, the assumption of a homogeneous steady state is not always compatible with the physics of this nonequilibrium system.
The linear response method applied to the onset of the instability suggests that a charge density wave (CDW), with well-defined wavelength (see Fig.~\ref{fig:densitywave}), sets in at later times.
To confirm this picture, we proceed now to the numerical solution of the nonlinear EOMs in the limit of small density.
In this case the computational complexity of the problem is greatly reduced, as each site is described by four numbers only.
The EOMs can be integrated quickly and reliably on very large lattices with $L \propto {\cal O}(10^{3})$ and for extremely long times $t \kappa \propto {\cal O}(10^{3})$.

\begin{figure}[t]
\begin{center}
\includegraphics[width=1.00\linewidth]{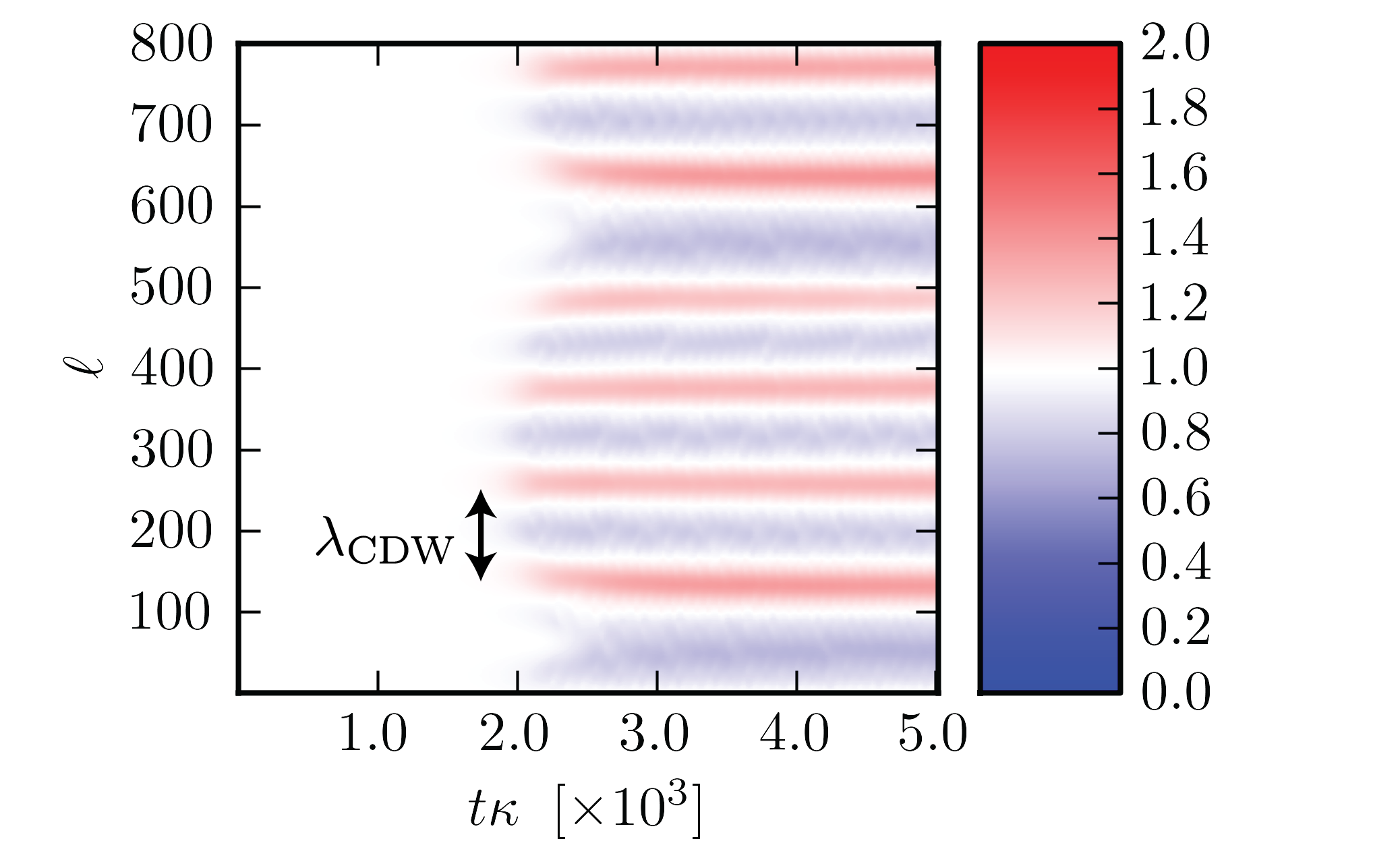}
\caption{Time-evolution of the rescaled density $n_{\ell}(t) / n$ of the system, according to the reduced EOMs for small density, for $L = 800$, $n = 0.1$, $J = 0.0$, and $U = 3.0$.
The system is initially prepared in a fully-condensed state and converges to a homogeneous steady state, which becomes unstable at $t\kappa \lesssim 2 \times 10^{3}$.
A persistent oscillation in the density profile is visible for later times. 
The double arrow shows the wavelength $\lambda_{\rm CDW} \simeq 120$ (in lattice units) of the instability as computed from the theory in Sec.~\ref{ssec:linresponse}. }
\label{fig:fullinstability}
\end{center}
\end{figure}

The spatial profile of the particle density is shown in Fig.~\ref{fig:fullinstability} for a typical time-evolution.
We clearly observe that a stationary CDW establishes at large times.
The wavelength $\lambda_{\rm CDW}$ of the periodic modulation is the same as the most unstable mode in the lowest-lying branch of the dissipative spectrum (the minima in the inset of Fig.~\ref{fig:spectrum}).
In other words, the fastest-growing mode in the continuum of unstable modes becomes dominant at large times \cite{BPSGD2009}, its wavelength is transmuted into the wavelength of the CDW, and, arguably, the remaining modes are eventually suppressed by nonlinear effects that are not described by the linear response approach used in Sec.~\ref{ssec:linresponse}.
This dynamically stable feature, with a relative density modulation $(n_{\rm max} - n_{\rm min}) / n \simeq 40\%$ in Fig.~\ref{fig:fullinstability}, gives an unambiguous, robust experimental signature of the fluctuation induced dynamical instability.
Typical optical lattice setups are smaller than $L = 800$ sites considered here, so it may be difficult in practice to obtain a periodic modulation if the most unstable mode has a very small momentum.
For a lattice with e.g.~$L \simeq 10^{2}$ sites, one would find a CDW where only a single wavelength fits into the optical lattice, and the measurement of the average density on the two halves of the lattice would be enough to detect the density fluctuation.
The reason to focus on larger lattices in our theoretical considerations is to exclude that we deal with a finite size effect.
Indeed, since only a small interval of momenta is dynamically unstable, it may happen for a lattice with $20$--$100$ sites that only the lowest momentum $q = 2\pi/L$ (in lattice units) in the discretized Brillouin zone is unstable (we reiterate that the dark state with $q = 0$ is always stable).
So, reducing to small lattices only, one cannot rule out that $\lambda_{\text{CDW}} = 1/q_{\text{CDW}} \propto L$, which would lead to a vanishing effect in the thermodynamic limit $L\to \infty$.
However, Fig.~\ref{fig:fullinstability} demonstrates that the spatial scale of the density modulation remains finite in the thermodynamic limit and coincides with the wavelength of the most unstable mode.

Intriguingly, the density profile presented in Fig.~\ref{fig:fullinstability} suggests that the system in the steady state spontaneously breaks the translational symmetry, which both Hamilton and Liouville operator present.
This adds to the breaking of the global phase symmetry due to the presence of the condensate.
A state where both symmetries are broken is often termed a ``supersolid'' although, in the equilibrium case, the breaking of the translational symmetry is stronger, with a spatial modulation on the order of the lattice constant \cite{SuSo}.
The spontaneous breaking of the translational symmetry discriminates the dynamical instability considered here from the dynamical instability suffered by an ensemble of interacting bosons moving with constant momentum through a lattice \cite{APDHL2005}.
In the latter case, if the momentum of the atomic ensemble is larger than an interaction-dependent threshold, the atomic current decays and the order in momentum space is lost.
The origin of this instability lies in the single-particle dispersion, that becomes negative in the outer regions of the Brillouin zone.
In our case, instead, the system occupies a momentum state $q = 0$ in which the single-particle dispersion is positive but the long-wavelength fluctuations of a collective (condensate) mode are destabilized by local fluctuations.
Moreover, in the case of Ref.~\cite{APDHL2005} the instability is purely classical because it stems from spatial fluctuations.
On the contrary, the local fluctuations that play the crucial role in our case are genuinely quantum, i.e.~temporal fluctuations.
Finally, while the order in momentum space is lost in the unstable atomic current, the external drive in our case allows the persistence of a (spontaneously) ordered pattern, in analogy to synergetic phenomena \cite{HakenBook}.

\section{Collapse and revival following a phase quench}
\label{sec:revival}
\begin{figure}[t]
\begin{center}
\includegraphics[width=1.00\linewidth]{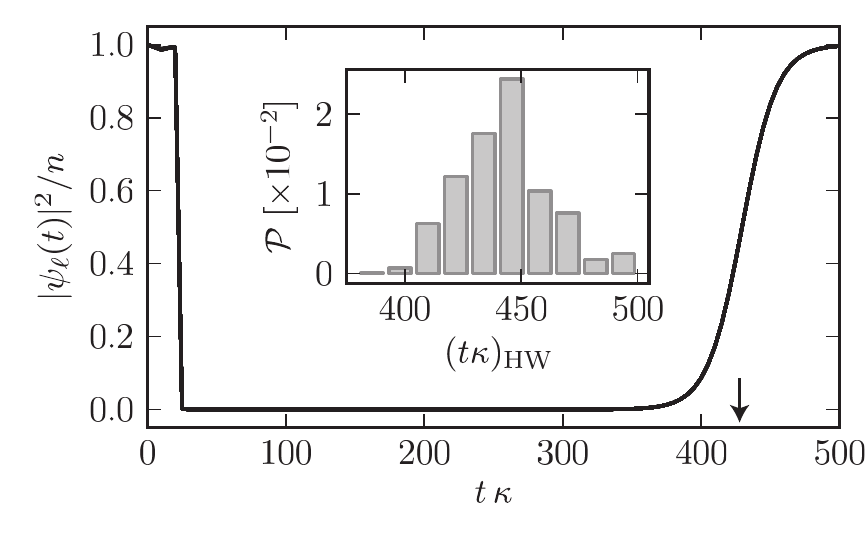}
\caption{The collapse and revival of the condensate fraction, evolved according to the EOM for small density, for $L = 32$, $n = 0.01$, and $J = U = 0.0$.
The initial state is a collection of local coherent states with random phases.
The phase $\phi$ is changed abruptly from $0$ to $3 \times 2 \pi / L$ at $(t \kappa)^{\ast} = 20.0$, after the system has reached the steady state.
The inset shows the probability density ${\cal P}$ for the time $(t\kappa)_{\rm HW}$ at which the revival of the condensate fraction on the first site of the lattice reaches $0.5$ (arrow in the main panel), obtained with $10^{3}$ evolutions with random initial phases.}
\label{fig:revival}
\end{center}
\end{figure}

In the previous sections we have considered the dynamics of the system in the presence of a stationary dissipative drive.
In the spirit of linear response theory, this has allowed us to study the dynamics of the system in terms of the instability of a collective mode.
In this section we focus on the response of the condensate to an abrupt change in the conditions of the drive.
The necessity to adapt to the widely different external conditions triggers a global response in the condensate and we observe intriguing nonlinear dissipative dynamics that can be understood in terms of collective variables.
Here we concentrate on the characteristics of this dynamics in the absence of unitary contributions to the time evolution, i.e.~$J=U=0$, with homogeneous density profile.
We first let the system converge to the steady state corresponding to $\phi = 0$, i.e.~a homogeneous condensate in the zero momentum mode, and then we change abruptly the phase $\phi$ in the Liouville operator at some finite time.
In the presence of a finite value of $\phi$, the dark state is a condensate with finite lattice momentum $-\phi$ and macroscopic inhomogeneous wavefunction $\psi_{\ell} \propto e^{i \phi \ell}$. 
We study the equilibration dynamics towards the new dark state.  
Our considerations apply to the mesoscopic dynamics \cite{NonlinDynExp} in periodic lattices, as implemented e.g.~in Ref.~\cite{HRMB2009} for purely coherent dynamics.
The dynamical effects that we identify are related to the finite size of these systems.

The time evolution is reported in Fig.~\ref{fig:revival} for a typical case.
The overall dynamics of the system consists of the rearrangement of the condensate fraction from the zero momentum of the initial state to the momentum $-\phi$ imprinted by the dissipative mechanism after the quench of the phase.
The rearrangement consists of an initial quick collapse of the condensate fraction and a subsequent revival with different momentum.
It is remarkable that the revival takes place after a substantial but well-defined time interval.
Indeed, in the particular case shown in the inset of Fig.~\ref{fig:revival}, the fluctuation of the time scale for the revival is of the order of $5\%$.
To understand the dynamical processes that take place during the transient of the rearrangement is the goal of the present section.
\begin{figure}[t]
\begin{center}
\includegraphics[width=1.00\linewidth]{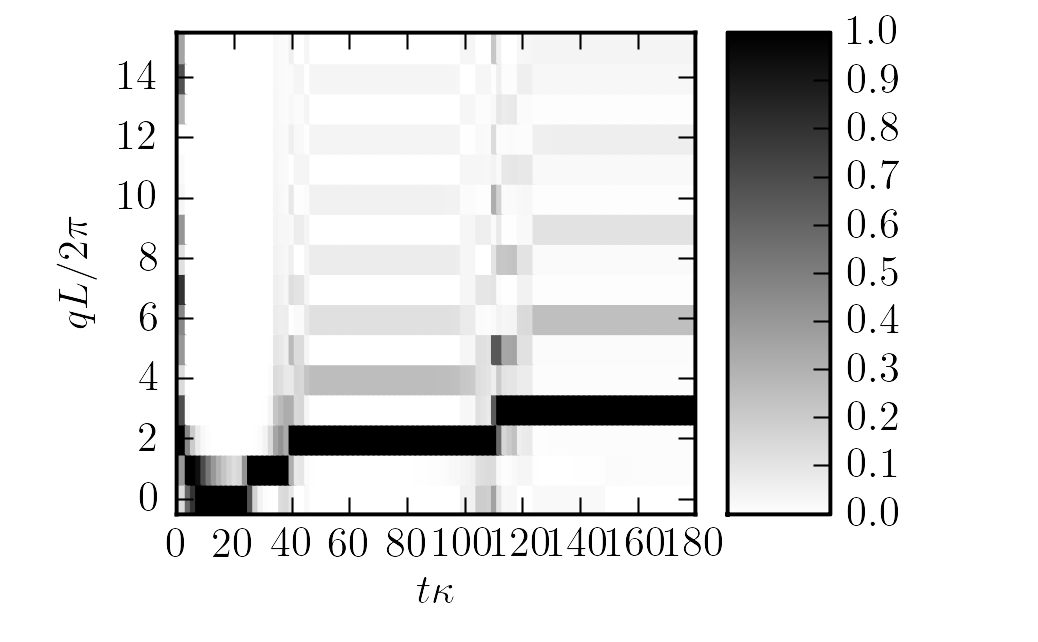}
\caption{The Fourier spectrum $|\theta_{q}|^{2}$ of the phases of the fluctuations $\langle \hat{b}_{\ell} \rangle = |\psi_{\ell}| e^{i \theta_{\ell}}$, for the same evolution shown in Fig.~\ref{fig:revival}.
The Fourier spectrum is normalized to its maximum at each distinct value of the time.
Since the distribution of phases $\theta_{\ell}$ is real, a symmetric peak appears in the negative part of the $q$ axis and is not shown in the plot. }
\label{fig:tunneling}
\end{center}
\end{figure}

Immediately after the quench, the condensate fraction decreases exponentially towards negligible values.
To describe the exponential decay we consider the EOMs (\ref{eq:corrhom}) and we keep only the terms that are linear in the powers of bosonic operators, that are dominant in the regime in which the correlations become negligible, obtaining
\begin{equation}
\partial_{t} \langle \hat{b}_{\ell} \rangle = - 4 \kappa [1 - \cos{(\phi)}] \langle \hat{b}_{\ell} \rangle~.
\end{equation}
We see that the initial state is depleted with a rate that increases as the final phase $\phi$ is quenched further from the initial value $\phi = 0$, where the initial state is stable.

To describe the revival we consider again the EOMs (\ref{eq:corrhom}) and it is convenient to use the moving frame of reference introduced in Sec.~\ref{sec:meanfield}.
The steady state of the system is a homogeneous condensate with respect to the correlations measured by the transformed operators $\tilde{b}_{\ell}$.
From the analysis of the instability performed in Sec.~\ref{sec:instability} we know that the effect of higher-order correlations is not necessarily negligible during the growth of a condensed mode.
Still, we can neglect the product of correlation functions, that are much smaller than the correlations themselves before the revival has taken place.
Taking the Fourier transform of the correlation functions with such prescription gives the coupled linear equations
\begin{equation}
\frac{1}{2 \kappa} \partial_{t}
\left ( \!\! \begin{array}{c} \langle \tilde{b} \rangle_{q} \\ \langle \tilde{n} \tilde{b} \rangle_{q} \end{array} \!\! \right ) =
\left ( \!\! \begin{array}{cc} 2(\cos{q} - 1) & 2 \cos{q} \\ 4 n & -4 \end{array} \!\! \right )
\left ( \!\! \begin{array}{c} \langle \tilde{b} \rangle_{q} \\ \langle \tilde{n} \tilde{b} \rangle_{q} \end{array} \!\! \right )~.
\end{equation}
Computing the eigenvalues of the matrix that defines the linear system we see that the growing mode that originates the revival can be located only at $q^{2} \lesssim n$, that in the original frame is a small interval centered around $-\phi$.
We note that reducing the equations of motion to the terms proportional to $\hat{b}_{\ell}$ only, as we did to describe the collapse of the condensate, is not appropriate in this case since the $q = -\phi$ mode would result marginally stable, but not unstable.
Hence, we identify a dynamical selection mechanism \cite{BL2006} that promotes the fluctuations in a certain interval of finite width.
It would be difficult to imagine a scenario in which a single mode is unstable, due to the continuity in $q$ of the equations of motion, which cannot give discontinuous results at any finite time.
The modes different from the dark state will be damped subsequently due to the action of the Liouvillian.

We turn now to the more interesting question, how the mode with lattice momentum $-\phi$ is initially populated during the time interval in which the condensate is depleted.
In other words, we are interested in the dynamics of the condensate fraction $\psi_{\ell} = \langle \hat{b}_{\ell} \rangle$, which exhibits tiny fluctuations around the fully-collapsed value $\psi_{\ell} = 0$ and which reorganizes at $\simeq (t \kappa)_{\rm HW}$ into the new reviving mode.
One naive expectation is that the phases $\theta_{\ell}$ of the coherent fluctuations $\psi_{\ell}(t) = |\psi_{\ell}(t)| e^{i \theta_{\ell}(t)}$ randomize and that the component $\theta_{q}$ with momentum $q = -\phi$ increases independently due to the process of dynamical selection outlined above.
On the contrary, it is visible from the Fourier spectra shown in Fig.~\ref{fig:tunneling} that the phases of the fluctuations exhibit a very organized dynamical phenomenon, featuring metastable steady states and rapid jumps into new configurations.
The Fourier spectrum has always a very clear peak that samples all the discrete points in the Brillouin zone, from the initial $q = 0$ to the final $q = -\phi$.
This contrasts indeed with the simpler organization of the phases that takes place starting from a random distribution, as can be seen in Fig.~\ref{fig:condensation3d} or for very short times in Fig.~\ref{fig:tunneling}.

In the following we argue that the momentum drift in discrete jumps shown in Fig.~\ref{fig:tunneling} can be understood as a cascade of macroscopic quantum tunneling (MQT) events between unstable steady states of the time-evolution.
Such picture is a qualitative modification of the well-known tunneling from a false vacuum into a global ground state that takes place in a system at quasi-equilibrium \cite{Coleman1977}.
To justify our picture, we study the dynamics of the phases $\theta_{\ell}$ only, given that the absolute value of the coherences $\psi_{\ell}$ is physically negligible.
The equation of motion for the phases reads, in general,
$i \partial_{t} \theta_{\ell}(t) = (2\psi_{\ell})^{-1} \partial_{t} \psi_{\ell} - (2\psi_{\ell}^{\ast})^{-1} \partial_{t} \psi_{\ell}^{\ast}$ and we specify this expression by keeping only the following particular term in the EOM for $\psi_{\ell}$
\begin{equation}\label{eq:psilaplacian}
\frac{1}{2\kappa}\partial_{t} \psi_{\ell} (t) = e^{i \phi} \psi_{\ell + 1}(t) - 2 \psi_{\ell}(t) + e^{-i \phi} \psi_{\ell - 1}~.
\end{equation}
The reason behind this choice is that, to the extent of identifying the metastable states of the phase dynamics, the linear approximation is enough.
Indeed, the linear term in the equations of motion is responsible for the exponential damping of the variables as we discussed in Sec.~\ref{sec:smalldens} in the context of critical dynamics, or at the beginning of the present section where the collapse of the condensate fraction is demonstrated.
The equation of motion for the phases then takes the form
\begin{equation}\label{eq:phasemotion}
\frac{1}{2\kappa} \partial_{t} \theta_{\ell}(t) = 2 \sin{\left ( \frac{\Delta_{\ell} \theta(t)}{2} \right )} \cos{\left ( \theta_{\ell + 1} - \theta_{\ell - 1} + 2 \phi \right )}~,
\end{equation}
where $\Delta_{\ell}$ is the discrete Laplace operator.
The structure of this equation is reminiscent of an overdamped string displaced by $\theta_{\ell}$ at the position $\ell$.
The kinetic term $\propto \partial_{t}^{2}$ is absent from the equation and the first derivative represents a friction term.
We conjecture that including the two neglected degrees of freedom (i.e.~the phases of $\langle \hat{n}_{\ell} \hat{b}_{\ell} \rangle$ and $\langle \hat{b}_{\ell}^{2} \rangle$) back into Eq.~(\ref{eq:psilaplacian}) would reinstate a kinetic term into Eq.~(\ref{eq:phasemotion}) and couple the phases to an effectively fluctuating field, although a more thoroughly designed approach (e.g.~within the Keldysh action formalism) would be necessary to maintain this assertion.

Following Ref.~\cite{DP2010}, let us parameterize the spatial dependence of the phases in terms of an ``instanton'' or phase slip configuration, $\theta_{\ell}(t) = Q(t) (\ell - 1) / (L - 1) + \delta\theta_{\ell}$, to distinguish the local dynamics $\delta \theta_{\ell}$ from the global variable $Q(t)$ describing the winding of the phases in the circular geometry.
It is important to notice that the difference $\delta \theta_{\ell + 1} - \delta \theta_{\ell}$ between phases on neighboring sites is defined in $[-\pi / 2, \pi / 2]$, to minimize the total variation of the phase profile.
With this convention for the difference between phases, the winding number reads $w = {\rm mod}[Q, 2\pi]$.
We first consider the case that the global variable is constant and study the dynamics of the spatial fluctuations, that move according to $(2\kappa)^{-1} \partial_{t} \delta\theta_{\ell} \simeq \cos{[\phi + \varphi]} \, \Delta_{\ell}\delta\theta$ with $\varphi = Q / (L-1)$. 
This is exactly the equation of an overdamped string, slightly displaced from its equilibrium configuration, hence all the fluctuations $\delta\theta_{\ell}$ converge to zero.
We consider now the dynamics of the global variable $Q(t)$.
If $0 < Q < \pi$, due to the boundary conditions we have that $\theta_{L+1} - \theta_{L} = -Q$.
The equation of motion for $Q(t)$ reads in this case $(2\kappa)^{-1} \partial_{t} Q(t) \simeq - \cos[\phi] \, Q$ and hence $Q$ decreases towards zero (we neglect terms that go to zero for large lattices).
If $\pi < Q < 2 \pi$, the boundary condition now is such that $\theta_{L+1} - \theta_{L} = 2 \pi - Q$, that entails $(2\kappa)^{-1} \partial_{t} [2\pi - Q(t)] \simeq - \cos[\phi] [2\pi - Q(t)]$ and hence $Q$ increases towards $2\pi$.
The same procedure applies to the other periodic replicas of the interval $[0, 2\pi)$ and shows that the global variable $Q(t)$ tends to the closer multiple of $2 \pi$.
In conclusion, we have shown that the state with $Q$ multiple of $2\pi$ (corresponding to a definite momentum in the distribution of the phases) is stable both with respect to local fluctuation of phases and to a global change of momentum.
We could hence think of a momentum state as a local minimum, whereto the system converges by an overdamped motion.

Having clarified the nature of the metastable configurations seen in Fig.~\ref{fig:tunneling}, we improve our conjecture on the effect of the terms that have been neglected in the equation of motion for $\theta_{\ell}$.
They necessarily provide a mechanism for the system to tunnel out of such minimum towards a nearby momentum configuration.
The picture of a sequence of tunneling events, each with a well-defined lifetime, explains the stability of the revival time over many configurations (see inset of Fig.~\ref{fig:revival}) and justifies a bell-shaped probability distribution for this quantity on the basis of the central limit theorem.
\begin{figure}[t]
\begin{center}
\includegraphics[width=1.00\linewidth]{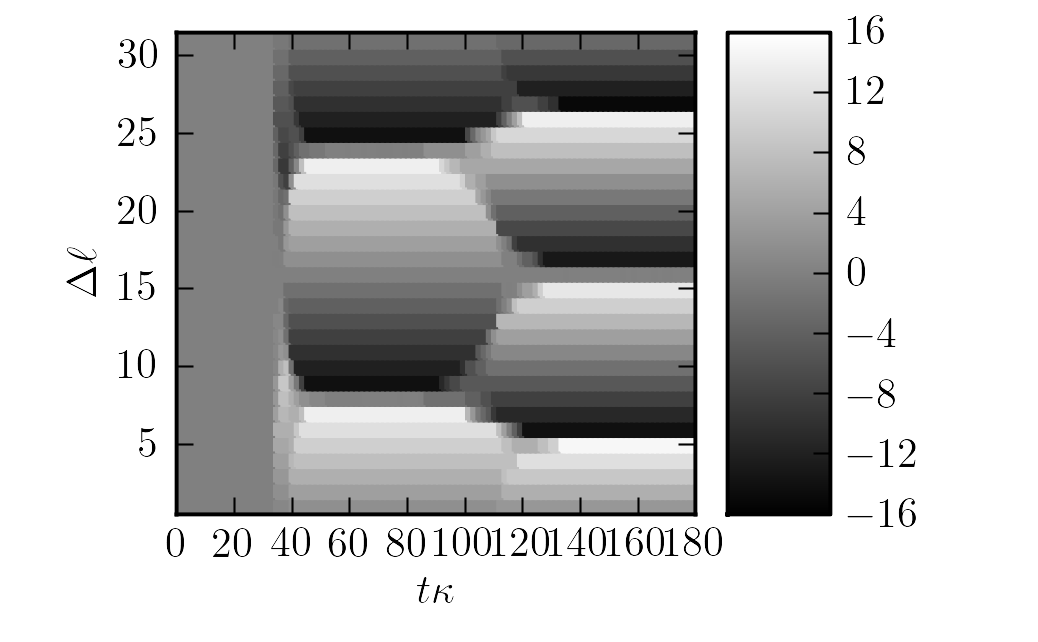}
\caption{The average phase $\bar{\theta}_{\Delta \ell}$, for the same evolution shown in Fig.~\ref{fig:revival}.
At each time, the winding number is given by the number of abrupt jumps in the average phase.
The locations of the phase jumps remain stationary in space, between the phase slip events, during which the winding number increases by one.}
\label{fig:slips}
\end{center}
\end{figure}

To further clarify the phase dynamics, we follow Ref.~\cite{DP2010} and compute the average phase difference $\bar{\theta}_{\Delta \ell} = \frac{1}{2 \pi} \sum_{\ell} {\rm Arg}\langle \hat{b}_{\ell}^{\dag} \hat{b}_{\ell + \Delta \ell} \rangle$ between sites at distance $\Delta \ell$.
For the initial configuration with zero momentum this quantity is zero everywhere.
We see that the initial phase configuration remains stable for a noteworthy time lag $20.0 < t \kappa \lesssim 35.0$ following the phase quench at $t \kappa \simeq 20.0$.
Then a fluctuation in the phases takes place and culminates in a phase slip at $t \kappa \simeq 40.0$, that increases the winding number of the phase distribution by one.
Another phase slip takes place at $t \kappa \gtrsim 40.0$ and the final one at $t \kappa \gtrsim 120.0$, corresponding to the abrupt jumps in the peaks of the Fourier spectrum shown in Fig.~\ref{fig:tunneling}.
We notice that the dynamics in momentum space takes place abruptly while it has a smooth appearance in real space: this discrepancy is reminiscent of the instanton description of the MQT phenomenon, in which the change of quantum numbers takes place instantaneously in imaginary time~\cite{DP2010}.
\begin{figure}[t]
\begin{center}
\includegraphics[width=1.00\linewidth]{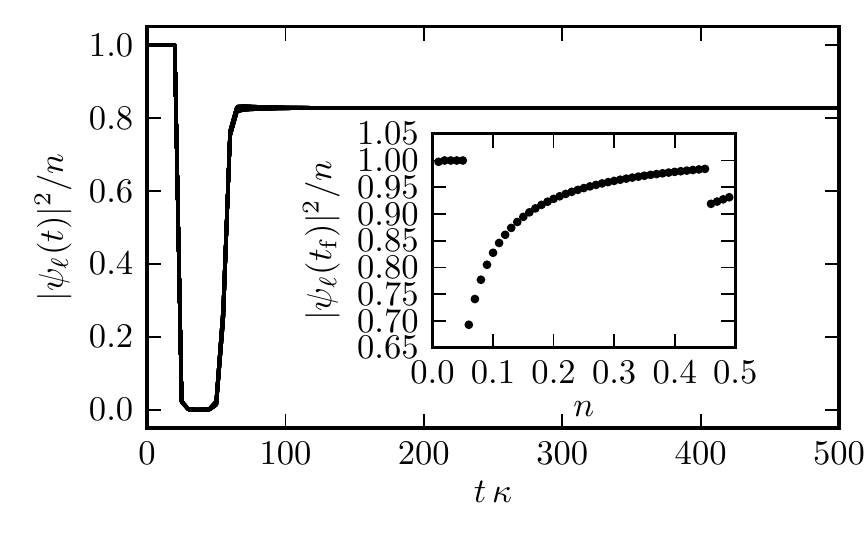}
\caption{Partial revival of the condensate fraction following a phase quench, obtained by integration of the reduced EOMs for small density, for $L = 32$, $n = 0.1$, and $J = U = 0.0$.
The phase $\phi$ is changed abruptly from $0$ to $3 \times 2\pi / L$ at $(t\kappa)^{\ast} = 20.0$.
For $t\kappa \lesssim 50.0$ the system stabilizes around the momentum $2 \times 2 \pi / L$ and develops a reduced metastable condensate fraction.
Nonetheless, convergence to the steady state with momentum $3 \times 2\pi / L$ is expected to take place at times larger than $T_{\rm f} \kappa = 500.0$.
The inset shows the condensate fraction at $T_{\rm f}$ as the density $n$ is varied.
For $n \lesssim 0.05$ the system reaches the steady state and the condensate revives fully (see also Fig.~\ref{fig:revival}). 
For $0.05 \lesssim n \lesssim 0.45$ the time-evolution is similar to the main panel, and the size of the condensate fraction depends strongly on the density.
For $0.45 \lesssim n$ the system develops a metastable condensate fraction with momentum $2\pi / L$. }
\label{fig:partial}
\end{center}
\end{figure}

We point out that the strong suppression of the condensate fraction, which takes place for very small densities, is not a necessary condition for the metastable behavior investigated above.
In Fig. 15 we report a typical time-evolution for larger densities (still well described by the reduced EOM for small density, see Fig.~\ref{fig:reduced}) in which the condensate fraction features a partial revival that persists as long as the true steady state is not reached.
While the explanation of the collapse of the condensate still holds, we do not have at present a dynamical model for the generation and persistence of a reduced condensate fraction.
As the density in the lattice is increased, the momentum of the metastable state changes and, at the transition point, a discontinuity in the fraction of revived condensate takes place.
To reproduce a curve like that shown in the inset of Fig.~\ref{fig:reduced} it is only necessary, in principle, to measure the condensate fraction at time $t \kappa \gtrsim 10^{2}$.
Hence, these features provide quite clear qualitative observable signatures of the underlying phase dynamics.

\section{Conclusions}
\label{sec:conclusions}

We have discussed various phenomena in driven-dissipative many-body boson systems governed by many-body master equations with a dissipative zero mode. 
This includes both a semi-quantitative characterization of the steady state phase diagram resulting from the competition of the driven-dissipative dynamics with a Hamiltonian, as well as a first characterization of mesoscopic nonlinear dissipative equilibration dynamics following phase quenches in the Liouville operator. 
For this purpose, we have developed a generalized Gutzwiller approach which accounts for density matrices corresponding to mixed states.
We have introduced a low density limit which exhibits all the qualitative features of the phase diagram and is amenable to an analytic discussion due to the decoupling of a finite subset of correlation functions. 

Several features are direct consequences of the fact that our system is far out of thermodynamic equilibrium. 
Most prominently, the phase transition in steady state shares features of a quantum phase transition in that it is interaction driven, and of a classical phase transition in that the ordered phase terminates in a thermal state, in stark contrast to dissipative quantum phase transitions in global thermodynamic equilibrium, where the constraint of zero temperature inhibits such behavior. 
Furthermore, in the limit of vanishing dissipation, the nonequilibrium phase diagram for the steady state does not connect smoothly to the phase diagram of the equilibrium Bose-Hubbard model -- this point is a manifestation of the non-commutativity of the limits $\kappa \to 0$ and $t \to \infty$. 
Another unconventional signature is the existence of a dynamical instability, which manifests itself in a spontaneous breaking of translational symmetry.
Finally, we have identified interesting nonlinear dissipative dynamics following a quench from one dark state to the other in mesoscopic systems, which hints at an interesting interplay of fluctuation and dissipation for effective macroscopic topological variables \cite{GardinerZollerBook,AltlandSimonsBook}.

We believe that the driven-dissipative many-body systems bear substantial potential for intriguing many-body physics far away from thermodynamic equilibrium. 
In particular, interesting questions concern the exploration of the critical behavior close to the phase transition and a more analytical understanding of the mesoscopic dissipative dynamics in terms of collective variables.

\acknowledgments
We acknowledge collaboration with Rosario Fazio and Andrea Micheli in the early stages of this work.
We acknowledge the European Union through IP AQUTE, and the Austrian Science Fund (FWF) through SFB F40 FOQUS.

\appendix

\section{Derivation of the mean field EOM}
\label{app:meanfield}

In this section we compute the right-hand side of Eq.~(\ref{eq:motiontrace}).
Let us start from the local interaction term in the Hamiltonian
\begin{eqnarray}
& & {\rm Tr}_{\neq \ell}[\sum_{\ell'} U \hat{n}_{\ell'}(\hat{n}_{\ell'} - 1),\rho] \nonumber \\
& = & {\rm Tr}_{\neq \ell} \sum_{\ell'} \bigotimes_{\ell''\neq \ell'} \rho_{\ell''} [U \hat{n}_{\ell'}(\hat{n}_{\ell'} - 1),\rho_{\ell'}]~.
\end{eqnarray}
In the summation above we separate the term with $\ell' = \ell$ from the others.
In the isolated term the trace applies only to the tensor product of density matrices and the commutator remains unaltered.
In the remaining terms, instead, the trace applies also to the commutator part and hence the contribution vanishes.
Hence, we obtain
\begin{equation}
{\rm Tr}_{\neq \ell}[\sum_{\ell'} U \hat{n}_{\ell'}(\hat{n}_{\ell'} - 1),\rho] = [U \hat{n}_{\ell}(\hat{n}_{\ell} - 1),\rho_{\ell
}]~.
\end{equation}
The contribution of the chemical potential, similarly, is the commutator with the term $- \mu \hat{n}_{\ell}$.
To treat a generic nonlocal term $O_{\ell\ell'}$ in the Hamiltonian or in the Lindblad structure we introduce the decomposition $O_{\ell\ell'} = \sum_{rs} \gamma_{\ell\ell'}^{rs} A_{\ell}^{r} \otimes B_{\ell'}^{s}$.
The hopping term $\hat{b}_{\ell}^{\dag}\hat{b}_{\ell'} + \hat{b}_{\ell'}^{\dag}\hat{b}_{\ell}$ on the $\ell$the site reads in this form
\begin{equation}
\gamma_{\ell\ell'}^{rs} = \delta_{rs},\quad A^{r}_{\ell} = (\hat{b}_{\ell},\hat{b}^{\dag}_{\ell})_{r},\quad B^{r}_{\ell} = (\hat{b}_{\ell}^{\dag},\hat{b}_{\ell})_{r}~,
\end{equation}
with $\ell$ and $\ell'$ nearest neighbors and $\delta_{rs}$ the Kronecker symbol.
The trace of the commutator gives
\begin{eqnarray}
& & {\rm Tr}_{\neq \ell}[\sum_{\ell'}\sum_{rs} \gamma_{\ell'\ell''}^{rs} A_{\ell'}^{r} \otimes B_{\ell''}^{s},\rho] \nonumber \\
& = & {\rm Tr}_{\neq \ell} \sum_{\ell'} \bigotimes_{\ell'''\neq\ell',\ell''} \rho_{\ell'''} [\sum_{rr'} \gamma_{\ell'\ell''}^{rr'} A_{\ell'}^{r} \otimes B_{\ell''}^{r'},\rho_{\ell'}\otimes \rho_{\ell''}] \nonumber \\
\end{eqnarray}
Breaking the sum over $\ell'$ as described above we obtain
\begin{equation}
{\rm Tr}_{\neq \ell}[-J\sum_{\ell\ell'}\hat{b}_{\ell}\hat{b}_{\ell'},\rho] = -J \sum_{\langle \ell' | \ell \rangle}[\hat{b}_{\ell} \langle \hat{b}_{\ell'} \rangle + \hat{b}_{\ell} \langle \hat{b}_{\ell'} \rangle,\rho_{\ell}]~,
\end{equation}
where $\langle \ell' | \ell \rangle$ indicates the sum over all the nearest neighbors $\ell'$ of $\ell$.
Finally, the contribution of the Hamiltonian is ${\rm Tr}_{\neq \ell}[\hat{\cal H},\rho(t)] = [\hat{h}_{\ell}(t), \rho_{\ell}(t)]$, with the effective locally-acting Hamiltonian
\begin{equation}\label{eq:localhamiltonian}
\hat{h}_{\ell} = -J\sum_{\langle \ell'| \ell\rangle}(\langle \hat{b}_{\ell'} \rangle \hat{b}_{\ell}^{\dag} +\langle \hat{b}_{\ell'}^{\dag} \rangle \hat{b}_{\ell}) + \frac{1}{2} U \hat{n}_{\ell} (\hat{n}_{\ell} - 1) - \mu \hat{n}_{\ell}~.
\end{equation}
In the Liouvillian term $\sum_{\langle ij \rangle}{\rm Tr}_{\neq\ell}[2c_{ij}\rho c_{ij}^{\dag} -c_{ij}^{\dag}c_{ij} \rho -\rho c_{ij}^{\dag} c_{ij}]$ in Eq.~(\ref{eq:motiontrace}) we insert the identity 
\begin{equation}
1 = 1 - \delta_{j \ell} - \delta_{i \ell} + \delta_{j \ell} +
\delta_{i \ell} + \delta_{i \ell} \delta_{j \ell} = (1 - \delta_{i
\ell}) (1 - \delta_{j \ell}) + \delta_{i \ell} + \delta_{j \ell}~,
\end{equation}
that holds since $j \neq i$ and so $\delta_{j \ell} \delta_{j \ell} = 0$ for every value of $\ell$.
The summation breaks into three parts
\begin{equation}
\left ( \sum_{\langle i j \rangle, i \neq \ell, i \neq \ell} +
\sum_{\langle i | \ell \rangle} \delta_{j \ell} + \sum_{\langle j |
\ell \rangle} \delta_{i \ell} \right ) {\rm Tr}_{\neq \ell}[\dots]~.
\end{equation}
The first term gives zero, since the expression ${\rm Tr}_{\neq \ell}[\dots]$ in this case is equivalent to $\rho_{\ell} {\rm Tr}[\dots]$ that vanishes for every couple of $i$ and $j$ because of the form of the Lindblad term.
Using that $\hat{c}_{\ell\ell'} = -\hat{c}_{\ell'\ell}$, the two remaining terms give
\begin{equation}\label{eq:traceliouv1}
2 \sum_{\langle \ell' | \ell \rangle} {\rm Tr}_{\neq \ell} [2 c_{\ell' \ell} \rho c_{\ell' \ell}^{\dag} - c_{\ell' \ell}^{\dag} c_{\ell' \ell} \rho -\rho c_{\ell' \ell}^{\dag} c_{\ell' \ell}]~.
\end{equation}
To proceed further, we assume a cubic square lattice geometry, with the coefficients $\phi_{\ell\ell'}$ that vanish on each plane orthogonal to the axial direction $\hat{x}$ and are equal to a constant $\phi$ along $\hat{x}$.
Then the phase difference between adjacent sites in the lattice is $\phi_{\ell\ell'} = \sigma \phi$, with $\sigma=\ell'-\ell$ along the $\hat{x}$ axis and $\sigma=0$ in the orthogonal plane.
The jump operators in the decomposition above read
\begin{equation}
\gamma_{\ell\ell'} = {\rm diag}(-1, -e^{i\sigma \phi}, e^{-i\sigma\phi}, 1)~,
\end{equation}
\begin{equation}\label{eq:liouvdecomp}
A_{\ell}^{r} = (1, \hat{b}_{\ell}^{\dag}, \hat{b}_{\ell}, \hat{n}_{\ell})_{r},\quad
B_{\ell}^{r} = (\hat{n}_{\ell}, \hat{b}_{\ell}, \hat{b}_{\ell}^{\dag}, 1)_{r}~.
\end{equation}
The contribution of a single neighbor in Eq.~(\ref{eq:traceliouv1}) is
\begin{equation}
\sum_{r,s} [2 A_{\ell}^{r} \rho_{\ell} A_{\ell}^{s\dag} - A_{\ell}^{s\dag} A_{\ell}^{r} \rho_{\ell} - \rho_{\ell} A_{\ell}^{s\dag} A_{\ell}^{r}] \Gamma_{\ell+1,\sigma}^{rs}~,
\end{equation}
with the matrix $\Gamma_{\ell, \sigma}^{r s}$ of correlation functions
\begin{equation}\label{eq:motionlattice}
\!\!\! \left ( \!\!\! \begin{array}{cccc} \langle
n_{\ell}^{2} \rangle_{\ell} & \langle b_{\ell}^{\dag}n_{\ell}
\rangle_{\ell} e^{-i\sigma\phi} & -\langle b_{\ell}n_{\ell}
\rangle_{\ell} e^{i\sigma\phi} & -\langle n_{\ell} \rangle_{\ell} \\
\langle n_{\ell}b_{\ell} \rangle_{\ell}e^{i\sigma\phi} & \langle
n_{\ell} \rangle_{\ell} & -\langle b_{\ell}^{2}
\rangle_{\ell}e^{i2\sigma\phi} & -\langle b_{\ell}
\rangle_{\ell}e^{i\sigma\phi} \\ -\langle n_{\ell}b_{\ell}^{\dag}
\rangle_{\ell}e^{-i\sigma\phi} & -\langle b_{\ell}^{2\dag}
\rangle_{\ell}e^{-i2\sigma\phi} & \langle n_{\ell} \rangle_{\ell}+1 &
\langle b_{\ell}^{\dag} \rangle_{\ell}e^{-i\sigma\phi} \\ -\langle
n_{\ell} \rangle_{\ell} & -\langle b_{\ell}^{\dag}
\rangle_{\ell}e^{-i\sigma\phi} & \langle b_{\ell}
\rangle_{\ell}e^{i\sigma\phi} & 1 \end{array} \!\!\! \right )_{\!rs}\!\!.
\end{equation}
The contribution of $\sigma=1$ and $\sigma=-1$ appears once, while $\sigma=0$ appears $z-2$ times, where $z$ is the coordination number of the lattice ($4$ for the two-dimensional square, $6$ for the three-dimensional cube).

\section{Linearization of the EOMs around the thermal state}
\label{app:linthermal}

We begin the analysis with the definition of the density matrix $\rho_{\ell} (t) = \rho_{\ell}^{\rm (th)} + \delta \rho_{\ell} (t)$ linearized around the thermal state and we keep only the first order in the entries of the perturbation $\delta \rho_{\ell}$.
The density matrix enters both the Hamiltonian $\hat{h}_{\ell} = \hat{h}_{\ell}^{(0)} + \delta \hat{h}_{\ell}$ and the Liouvillian ${\cal L}_{\ell} = {\cal L}_{\ell}^{(0)} + \delta {\cal L}_{\ell}$ in the computation of the correlations.
To the first order, the EOM for the perturbation matrix reads
\begin{equation}\label{eq:linliouvmf}
\partial_{t} \delta \rho_{\ell} (t) + i[\hat{h}_{\ell}^{(0)}, \delta
\rho_{\ell} (t)] - {\cal L}_{\ell}^{(0)}[\delta \rho_{\ell} (t)]  = -
i[\delta \hat{h}_{\ell}, \rho_{\ell}^{\rm (th)}] + \delta {\cal L}_{\ell}
[\rho_{\ell}^{\rm (th)}]~,
\end{equation}
where the left-hand side is linear in the perturbation matrix, while the right-hand side acts as an effective driving. 
The right-hand side depends on the perturbation density matrices $\delta \rho_{\ell \pm 1}$ on the neighboring sites of $\ell$ via the correlation functions that enter the Hamiltonian and the Liouvillian.
The terms in Eq.~(\ref{eq:linliouvmf}) read
\begin{eqnarray}
\hat{h}_{\ell}^{(0)} & = & \frac{U}{2} \hat{b}_{\ell}^{\dag} \hat{b}_{\ell}^{\dag} \hat{b}_{\ell} \hat{b}_{\ell}~, \\
\delta \hat{h}_{\ell}(t) & = & -J \hat{b}_{\ell} \sum_{\sigma} \delta \langle \hat{b}_{\ell + \sigma}^{\dag} \rangle -J \hat{b}_{\ell}^{\dag} \sum_{\sigma} \delta \langle \hat{b}_{\ell + \sigma} \rangle ~, \nonumber \\
{\cal L}_{\ell}^{(0)}[\rho] & = & \kappa \sum_{rs} [2 \hat{A}_{\ell}^{r} \rho \hat{A}_{\ell}^{s \dag} -\hat{A}_{\ell}^{s \dag} \hat{A}_{\ell}^{r} \rho - \rho \hat{A}_{\ell}^{s \dag} \hat{A}_{\ell}^{r} ] \sum_{\sigma = \pm 1} \Gamma_{\ell + \sigma}^{(0)\, rs}~, \nonumber \\
\delta {\cal L}_{\ell}[\rho] & = & \kappa \sum_{rs} [2 \hat{A}_{\ell}^{r} \rho \hat{A}_{\ell}^{s \dag} -\hat{A}_{\ell}^{s \dag} \hat{A}_{\ell}^{r} \rho - \rho \hat{A}_{\ell}^{s \dag} \hat{A}_{\ell}^{r} ] \sum_{\sigma = \pm 1} \delta \Gamma_{\ell + \sigma}^{rs}~, \nonumber
\end{eqnarray}
where we have reserved the possibility for the perturbation to be inhomogeneous in real space.
The unperturbed matrix of the zeroth-order correlations, computed on the thermal state, reads
\begin{equation}
\Gamma_{\ell}^{(0)} = \left ( 
\begin{array}{cccc} 
2 n (n + 1/2) & 0 & 0 & - n \\
0 & n & 0 & 0 \\
0 & 0 & n + 1 & 0 \\
- n & 0 & 0 & 1
\end{array} \right )~.
\end{equation}
The matrix of the first-order correlations is
\begin{equation}
\delta \Gamma_{\ell} = \left ( \begin{array}{cccc}
\delta \langle \hat{n}_{\ell}^{2} \rangle & \delta \langle \hat{n}_{\ell} \hat{b}_{\ell} \rangle^{\ast} & - \delta \langle \hat{b}_{\ell} \hat{n}_{\ell} \rangle & - \delta \langle \hat{n}_{\ell} \rangle \\ 
\delta \langle \hat{n}_{\ell} \hat{b}_{\ell} \rangle & \delta \langle \hat{n}_{\ell} \rangle & -\delta \langle \hat{b}_{\ell}^{2} \rangle & - \delta \langle \hat{b}_{\ell} \rangle \\
-\delta \langle \hat{b}_{\ell} \hat{n}_{\ell} \rangle^{\ast} & - \delta \langle \hat{b}_{\ell}^{2} \rangle^{\ast} & \delta \langle \hat{n}_{\ell} \rangle & \delta \langle \hat{b}_{\ell} \rangle^{\ast} \\
- \delta \langle \hat{n}_{\ell} \rangle & - \delta \langle \hat{b}_{\ell} \rangle^{\ast} & \delta \langle \hat{b}_{\ell} \rangle & 0 
\end{array} \right
)~,
\end{equation}
with the averages $\delta \langle \dots \rangle = {\rm Tr}[\dots \, \delta \rho_{\ell}] $.
We verified with the numerical integrations of the nonlinear EOM (\ref{eq:motiontrace}) that $\delta \langle \hat{b}_{\ell}^{2} \rangle$ is much smaller than the other correlation functions and hence can be neglected.

The structure of the EOM for $\delta\rho_{\ell}$ is determined by the right-hand side of Eq.~(\ref{eq:linliouvmf}), that is tridiagonal in the Fock space.
It follows that only the entries of $\delta \rho_{\ell}$ on the upper, central, and lower diagonal need to be taken different from zero. 
Moreover, the equations for the entries in each diagonal constitute a closed system, so that we have only to consider the variables $x_{\nu} = \delta \rho_{\ell; \nu, \nu-1}$, where $\nu \ge 0$ is the index in the Fock space.

We derive a set of rules to project the EOM onto the lower diagonal.
For tridiagonal matrices (such as $\delta \rho_{\ell}$) it holds that
\begin{equation}
\begin{array}{cccccc}
\rho & \to & x_{\nu} &  \hat{b}^{\dag} \hat{b} \rho & \to & \nu x_{\nu} \\
\rho \hat{b}^{\dag} \hat{b} & \to & (\nu - 1) x_{\nu} & \hat{b}^{\dag} \hat{b} \hat{b}^{\dag} \hat{b} \rho & \to & \nu^{2} x_{\nu} \\
\rho \hat{b}^{\dag} \hat{b} \hat{b}^{\dag} \hat{b} & \to & (\nu - 1)^{2} x_{\nu} & \hat{b}^{\dag} \hat{b} \rho \hat{b}^{\dag} \hat{b} & \to & \nu (\nu - 1) x_{\nu} \\
\hat{b} \rho \hat{b}^{\dag} & \to & \sqrt{\nu (\nu + 1)} x_{\nu + 1} & \hat{b}^{\dag} \rho \hat{b} & \to & \sqrt{\nu (\nu - 1)} x_{\nu - 1}~,
\end{array}
\end{equation}
and for diagonal matrices (such as $\rho_{\ell}^{\rm (th)}$) it also holds that
\begin{equation}
\begin{array}{cccccc}
\rho \hat{b} & \to & 0 & \rho \hat{b}^{\dag} & \to & \sqrt{\nu} x_{\nu} \\
\hat{b} \rho & \to & 0 & \hat{b}^{\dag} \rho & \to & \sqrt{\nu} x_{\nu - 1} \\
\hat{b}^{\dag} \rho \hat{b}^{\dag} \hat{b} & \to & \sqrt{\nu} (\nu - 1) x_{\nu - 1} & \hat{b} \rho \hat{b}^{\dag} \hat{b} & \to & 0 \\
\hat{b}^{\dag} \hat{b} \rho \hat{b}^{\dag} & \to & \sqrt{\nu} \nu x_{\nu} & \hat{b}^{\dag} \hat{b} \rho \hat{b} & \to & 0 \\
\rho \hat{b}^{\dag} \hat{b} \hat{b}^{\dag} & \to & \sqrt{\nu} \nu x_{\nu} & \hat{b}^{\dag} \hat{b}^{\dag} \hat{b} \rho & \to & \sqrt{\nu} (\nu - 1) x_{\nu - 1} \\
\rho \hat{b}^{\dag} \hat{b} \hat{b} & \to & 0 & \hat{b} \hat{b}^{\dag} \hat{b} \rho & \to & 0~.
\end{array}
\end{equation}
We apply these rules to the EOM (\ref{eq:linliouvmf}) for the perturbation matrix and we obtain the EOM for the variables $x_{\nu}$
\begin{eqnarray}\label{eq:eqmotionsubdiag}
\frac{1}{2 \kappa} \partial_{t} x_{\nu} 
& = & \frac{1}{2} [-\nu(4 + 8 n + i U / \kappa) + i U / \kappa ] x_{\nu} \nonumber \\
& + & 2 n \sqrt{\nu (\nu - 1)} x_{\nu - 1} + 2 (n + 1) \sqrt{\nu (\nu + 1)} x_{\nu + 1} \nonumber \\
& + & \sqrt{\nu} (\delta \langle \hat{b}_{\ell - 1} \rangle + \delta \langle \hat{b}_{\ell + 1} \rangle) (\rho_{\ell; \nu - 1, \nu - 1}^{\rm (th)} + \rho_{\ell; \nu, \nu}^{\rm (th)}) \nonumber \\
& + & \frac{1}{2}\sqrt{\nu} [-\delta \langle \hat{b}_{\ell - 1} \rangle ( 2\nu -i J / \kappa) \nonumber \\
& - & \delta \langle \hat{b}_{\ell + 1} \rangle (2 \nu -i J / \kappa) \nonumber \\
& + & 2 (\delta \langle \hat{b}_{\ell - 1} \hat{n}_{\ell - 1} \rangle + \delta \langle \hat{b}_{\ell + 1} \hat{n}_{\ell + 1} \rangle)] \nonumber \\
& \times & (\rho_{\ell; \nu - 1, \nu - 1}^{\rm (th)} - \rho_{\ell; \nu, \nu}^{\rm (th)})~.
\end{eqnarray}

We see that in the EOMs the subdiagonal entries of the perturbation matrix on the site $\ell$ are coupled to some correlation functions computed on the neighboring sites.
To allow for any possible spatial configuration of the correlations makes the problem analytically unmanageable.
Moreover, to compute the linearized equations on a finite number of lattice sites induces a discretization of the Brillouin zone such that the numerical effort to solve the linear equations scales quadratically with the size of the lattice.
At this point it is necessary to introduce some physical hypothesis on the form of the instability, which allows us to consider only a finite number of degrees of freedom and to easily scan the Brillouin zone of the lattice.
Here we make the hypothesis
\begin{eqnarray}\label{eq:ansatzlatticeinstab}
\delta \langle \hat{b}_{\ell'} \rangle & = & \delta \langle \hat{b}_{\ell} \rangle e^{- i \phi_{0}(\ell' - \ell)}~, \nonumber \\
\delta \langle \hat{b}_{\ell'} \hat{n}_{\ell'} \rangle & = & \delta \langle \hat{b}_{\ell} \hat{n}_{\ell} \rangle e^{- i \phi_{0}(\ell' - \ell)}~.
\end{eqnarray}

Using Eq.~(\ref{eq:ansatzlatticeinstab}) we reduce the set of variables in the EOM (\ref{eq:eqmotionsubdiag}) to the sole $\{x_{\nu}\}$.
The EOM is rewritten in the form $\partial_{t} x_{\nu} = \sum_{\nu'} M_{\nu,\nu'} x_{\nu'}$, where
\begin{eqnarray}\label{eq:instabEq}
M_{\nu,\nu'}/(2\kappa)
& = & 2 n \sqrt{\nu (\nu - 1)} \delta_{\nu', \nu - 1} + \frac{1}{2} [-\nu (4 + 8 n + i \frac{U}{\kappa}) \nonumber \\
& + & i \frac{U}{\kappa}] \delta_{\nu,\nu'} + 2 (n + 1) \sqrt{\nu (\nu + 1)} \delta_{\nu', \nu + 1} \nonumber \\
& + & \sqrt{\nu \nu'} [ \cos{(\phi_{0})} 2 (\nu' -  \nu) + i \frac{J}{\kappa} \cos{\phi_{0}} ] \nonumber \\
& \times & (\rho_{\ell; \nu - 1, \nu - 1}^{(0)} - \rho_{\ell; \nu, \nu}^{(0)}) \nonumber \\
& + & 2 \cos{(\phi_{0})} \sqrt{\nu' \nu} (\rho_{\ell; \nu - 1, \nu - 1}^{(0)} + \rho_{\ell; \nu, \nu}^{(0)})~,
\end{eqnarray}
as can be verified with the formal substitutions 
\begin{equation}\label{eq:corrfuncspatial}
x_{\nu} \to \delta_{\nu,\nu'},\quad
\delta \langle \hat{b}_{\ell} \rangle \to \sqrt{\nu'},\quad
\delta \langle \hat{b}_{\ell} \hat{n}_{\ell} \rangle \to \nu' \sqrt{\nu'}~.
\end{equation}
The general solution of the EOM is given by $x_{\nu}\sim e^{\lambda t}$, with $\lambda$ an eigenvalue of the matrix $M$.
If ${\rm Re}[\lambda] < 0$ for any $\lambda$, the perturbation $\delta \rho_{\ell}$ damps out and the thermal state is stable.
On the contrary, if there is at least one eigenvalue $\lambda$ with a positive real part $\gamma = {\rm Re}[\lambda]$ the thermal state is unstable.
The corresponding eigenvector is the unstable mode in the channel labelled by $\phi_{0}$.
To diagonalize the matrix $M$ it is necessary to introduce a truncation $\nu_{\rm max}$ in the set of variables $\{x_{\nu}\}$, corresponding to a truncation in the Fock space of the system.
We use $\nu_{\rm max} = 100$ and we have verified in some typical cases that the higher part of the spectrum remains unchanged when $\nu_{\rm max}$ is increased.

\section{Linearization of the EOMs around a generic mean field steady state}
\label{app:lingeneral}

In this section we outline the computation of the stability spectrum of a generic mean field homogeneous steady state with local density matrix $\rho_{\infty}$.
This method is used in Sec.~\ref{ssec:stability} to compute the phase border between the stable and the unstable condensate region.
The mean field EOMs (\ref{eq:motiontrace}) can be rewritten in the form $\partial_{t}\rho_{\ell}(t) = Q[\rho_{\ell}(t),\rho_{\ell \pm 1}(t)]$, with $Q$ a bilinear operator.
Let us characterize now the action of the operator $Q$ on the steady-state $\rho_{\ell}(t_{\infty})$ of the system.
We require the steady state to be such that $\partial_{t} |\rho_{\ell;\nu,\nu'}(t_{\infty})| = 0$, where $\nu$ and $\nu'$ are indices in the Fock space of the $\ell$th site.
It is easy to show that a sufficient and necessary condition for this to happen is $\partial_{t} \rho_{\ell;\nu,\nu'}(t_{\infty}) = - i \Omega_{\nu\nu'} \rho_{\ell;\nu,\nu'}$, with $\Omega_{\nu\nu'} \in \mathbb{R}$.
We also require that the absolute value of a generic correlation function $\langle \hat{b}_{\ell}^{\dag p} \hat{b}_{\ell}^{q} \rangle$ is constant in time, and hence $\partial_{t} \langle \hat{b}_{\ell}^{\dag p} \hat{b}_{\ell}^{q} \rangle = -i \Lambda_{pq} \langle \hat{b}_{\ell}^{\dag p} \hat{b}_{\ell}^{q} \rangle $.
The stationarity of the absolute values of the correlation functions can be easily checked on the numerical solution of the equations of motion.

Still, we cannot forbid a residual phase precession both in the off-diagonal elements of the density matrix and in the correlation functions.
The first point raises a problem for the linearization of the equations of motion because we do not obtain numerically a state that nullifies exactly the bilinear operator $Q$, i.e.~we do not have a good starting point for the instability analysis.
It turns out that a chemical potential $\mu$ can always be fixed into the operator $Q$ (that we denote $Q_{\mu}$ after the choice is made) such that $Q_{\mu}[\rho_{\ell}(t_{\infty}),\rho_{\ell\pm 1}(t_{\infty})] = 0$ exactly.
To explain why this is the case, we consider the explicit form of the correlation functions $\langle \hat{b}_{\ell}^{\dag p} \hat{b}_{\ell}^{q} \rangle = \sum_{\nu} f(q,\nu + q) f(p, \nu) \rho_{\ell;\nu + q, \nu + p}$, with $f(p,\nu) = \langle \nu + p | \hat{b}^{\dag p}| \nu \rangle$.
From the conditions above it follows that $\Omega_{\nu+p,\nu+q} = \Lambda_{pq}$ and hence $\Omega$ can only depend on the difference of its arguments and be of the form $\Omega_{p - q}$.
Now we may argue that $\Omega$ should have the linear form $\Omega\times(p-q)$ since we want the phase frequency of $\langle \hat{b}^{2}_{\ell} \rangle$ be twice that of $\langle \hat{b}_{\ell} \rangle$ in the condensate limit.
Finally, $\Omega$ should be a linear function of time if we require that we can remove the phase precession by a frame transformation of the basis vectors.
The result is that, in the steady state, the residual time-evolution of the density matrix is of the form $\partial_{t}\rho_{\ell;\nu,\nu'} = -i \omega (\nu - \nu') \rho_{\ell;\nu,\nu'}$.
This is precisely the contribution to the equation of motion produced by a term $\omega \hat{n}$ in the Hamiltonian and can thus be eliminated by the appropriate choice of a chemical potential $\mu = -\omega$.

With such choice of chemical potential, the equation of motion for the fluctuation $\delta \rho_{\ell}(t) = \rho_{\ell}(t) - \rho_{\infty}$ of the density matrix reads $\partial_{t}\delta\rho_{\ell}(t) = Q_{\mu}[\rho_{\infty}, \delta \rho_{\ell \pm 1}(t)] + Q_{\mu}[\delta \rho_{\ell}(t), \rho_{\infty}]$.
Then, as we point out in Sec.~\ref{ssec:stability} as well, some hypotheses on the spatial dependence of the unstable mode has to be made to relate the density matrices on different sites as $\delta\rho_{\ell \pm 1} = T^{(\pm)}[\delta\rho_{\ell}, \phi_{0}]$, where we restrict to a one-dimensional geometry.
According to the results of the numerical solution of the EOM (\ref{eq:motiontrace}), here we consider the stability of the modes defined by $[T^{(\pm)}[\delta\rho_{\ell}, \phi_{0}]]_{\nu\nu'} = \cos{(\phi_{0})} [\rho_{\ell}]_{\nu\nu'}$, that correspond to the generation of a standing wave in the lattice with wavelength $2\pi / \phi_{0}$.
The EOM for the fluctuation matrix then reads $\delta \rho_{\ell}(t) = L[\delta \rho_{\ell},\phi_{0}]$ with the linear operator $L$ given by $Q_{\mu}[\rho_{\ell}(t_{0}), \cdot] \circ (T^{(+)}[\cdot, \phi_{0}] + T^{(-)}[\cdot, \phi_{0}]) + Q_{\mu}[\cdot, \rho_{\ell}(t_{0})]$.
The existence of a positive real eigenvalue of $L[\cdot,\phi_{0}]$ means that the steady state $\rho_{\infty}$ is unstable in the particular mode labeled by $\phi_{0}$.

\section{EOM of the correlation functions}
\label{app:eqmotion}

In this section we provide the EOMs of the time-dependent correlation functions $c_{p,q,\ell} \equiv \langle \hat{b}_{\ell}^{\dag p} \hat{b}_{\ell}^{q} \rangle$ that follow from the mean field EOM (\ref{eq:motiontrace}) for the density matrix.
For definiteness, in the following we assume a one-dimensional geometry so that the neighbors $\langle \ell' | \ell \rangle$ of the sites $\ell$ have indices $\ell+1$ and $\ell - 1$.
The EOM of a generic correlation function is $\partial_{t}\langle \hat{b}_{\ell}^{\dag p} \hat{b}_{\ell}^{q} \rangle = {\rm Tr}[ \hat{b}_{\ell}^{\dag p} \hat{b}_{\ell}^{q} \partial_{t}\rho_{\ell}(t) ]$.
Substituting the time derivative of the density matrix $\rho_{\ell}$ one obtains products of averages of operators on the site $\ell$ by the mean fields (\ref{eq:corrset}) on the neighboring sites.
Normal ordering the operators acting on the site $\ell$, one obtains the following expression, which is linear in the correlation functions on the site $\ell$ but contains products between correlation functions on neighboring sites
\begin{widetext}
\begin{eqnarray}\label{eq:motionanycorr}
\partial_{t} c_{p,q,\ell} & = &
+ \left [ i (U / 2) ( p(p-1) - q(q-1) ) - 2 \kappa ( (p - q)^{2} + p + q )-i \mu (p - q) \right ] c_{p,q,\ell} \nonumber \\
& & +i U (p-q) c_{p+1,q+1,\ell} + 2 \kappa p q (c_{1,1,\ell-1}+ c_{1,1,\ell+1}) c_{p-1,q-1,\ell} \nonumber \\
& & + \left [
- i J p (c_{1,0,\ell-1} + c_{1,0,\ell+1})
+ 2 \kappa p (1 - q) (e^{i \phi } c_{1,0,\ell-1} + e^{-i \phi } c_{1,0,\ell+1})
+ 2 \kappa p  (e^{i \phi } c_{2,1,\ell-1} + e^{-i \phi } c_{2,1,\ell+1})
\right ] c_{p-1,q,\ell}  \nonumber \\
& & + \left [
i J q (c_{0,1,\ell-1} + c_{0,1,\ell+1})
+ 2 \kappa q (1 - p) (e^{-i \phi } c_{0,1,\ell-1} + e^{i \phi } c_{0,1,\ell+1})
+ 2 \kappa q (e^{-i \phi } c_{1,2,\ell-1} + e^{i \phi } c_{1,2,\ell+1} )
\right ] c_{p,q-1,\ell} \nonumber \\
& & - 2 q \kappa (e^{i \phi } c_{1,0,\ell-1} + e^{-i \phi } c_{1,0,\ell+1}) c_{p,q+1,\ell} 
- 2 p \kappa (e^{-i \phi } c_{0,1,\ell-1} + e^{i \phi } c_{0,1,\ell+1}) c_{p+1,q,\ell} \nonumber \\
& & + \kappa p(p-1) (e^{2 i \phi } c_{2,0,\ell-1} + e^{-2 i \phi } c_{2,0,\ell+1}) c_{p-2,q,\ell}
+ \kappa q(q-1) (e^{-2 i \phi } c_{0,2,\ell-1} + e^{2 i \phi } c_{0,2,\ell+1}) c_{p,q-2,\ell}~,
\end{eqnarray}
\end{widetext}
where it is understood that $c_{p,q,\ell} = 0$ whenever $p,q < 0$ and $c_{0,0,\ell} = 1$ by the unitarity of the trace of $\rho_{\ell}$.
The above expression shows that the EOMs of the correlation functions are organized as an infinite hierarchy, since $c_{p,q,\ell}$ is also coupled to $c_{p+1,p+1,\ell}$ (due to the local interaction) and to $c_{p,q+1,\ell}$, $c_{p+1,q,\ell}$ (due to the dissipation).
Hence we do not find a subset of EOMs that decouples exactly from the others and that can be solved in a finite number of steps.

In Sec.~(\ref{sec:smalldens}) we use the different set of ``connected'' correlation functions $\langle \delta \hat{b}_{\ell}^{\dag p} \delta \hat{b}_{\ell}^{q} \rangle$, built with the fluctuation operator $\delta \hat{b}_{\ell} = \hat{b}_{\ell} - \psi_{\ell}$.
We remark that we always work in the Schr\"odinger picture here: the operators $\delta \hat{b}_{\ell}$ are constant in time and depend parametrically on the choice of the quantity $\psi_{\ell}$.
This set is totally equivalent from a mathematical point of view, but is motivated from a physical perspective if we interpret $\psi_{\ell}$ as the order parameter of a Bose-Einstein condensate and (\ref{eq:corrset}) holds.
The new set of correlation functions hence describes the fluctuations about the condensate and, by construction $\langle \delta \hat{b}_{\ell} \rangle = 0$.
We reiterate that the vanishing of the first-order correlation takes place only when the average is taken with respect to some density matrix $\rho_{\ell}(t_{0})$ such that (\ref{eq:corrset}) is valid.
Once we let the density matrix evolve in time, as we do in Sec.~\ref{sec:instability}, the operators $\delta \hat{b}_{\ell}$ do not change but the average ${\rm Tr}[\delta \hat{b}_{\ell} \rho_{\ell} (t)] = {\rm Tr}[\hat{b}_{\ell} \rho_{\ell} (t)] - \psi_{\ell}$ is in general different from zero.

To obtain the EOM for the correlation functions $\delta \hat{b}_{\ell}$ we expand the product of $\hat{b}_{\ell}$ operators and $\psi_{\ell}$ $c$-numbers in the average, obtaining a sum of correlation functions $c_{p,q,\ell}$.
Then we substitute the EOM (\ref{eq:motionanycorr}) of each $c_{p,q,\ell}$ and again we expand the products of operators in the averages in terms of $\delta \hat{b}_{\ell}$ and $\psi_{\ell}$.
This approach does not yield the general EOM, analogous to (\ref{eq:motionanycorr}), because each choice of $p$, $q$ produces a different set of correlations, but can be applied straightforwardly.
The EOMs for the connected correlation functions are organized as an infinite hierarchy as well, and cannot be reduced analytically to a closed subset.


\begin{thebibliography}{77}
\bibitem{footnote2}
We note that the thermal equation (\ref{eq:thevol}) admits solutions which are quasi-local in Fock space, $\rho_{0,0} =  \frac{N+ 1}{2 N+1}, \quad \rho_{1,1} =  \frac{N}{2 N+1}, \quad \rho_{n,n} =  0 \quad \mathrm{for} \quad n>1$, but we do not observe that they are reached in a time evolution starting from an initial coherent state.
\bibitem{footnote3}
Note the subtlety that, while a Bogoliubov treatment with factorized higher order correlation functions yields $\mu = Un_0$ ($n_0$ the condensate density, cf.~Sec.~\ref{ssec:phases}), the solution of our coupled low density equations in steady state yields the result $\mu = Un$ with full average density.
\end{thebibliography}
\end{document}